\PassOptionsToPackage{dvipsnames}{xcolor}	
\documentclass[12pt,letter]{article}

\usepackage[T1]{fontenc}
\usepackage{lmodern}
\usepackage{setspace}
\usepackage{etoolbox}
\makeatletter
\patchcmd{\appendix}{\@Alph}{\@Roman}{}{}
\makeatother

\usepackage{amsmath}
\usepackage{amssymb}
\usepackage{amsthm}
\usepackage{mathtools}
\usepackage{mathrsfs}
\usepackage{breqn}
\usepackage{bigints}
\usepackage{bm}
\usepackage{bbm}

\usepackage[margin=1 in]{geometry}
\usepackage{enumerate}
\usepackage[shortlabels]{enumitem} 
\usepackage{xcolor}
\usepackage[colorlinks=true, urlcolor=blue,  linkcolor=blue, citecolor=blue]{hyperref}
\usepackage{float}
\usepackage{indentfirst}
\usepackage{caption}
\usepackage{fancyhdr}
\usepackage{tcolorbox}
\usepackage{multirow,array}
\usepackage{booktabs}
\usepackage[noabbrev,capitalise]{cleveref}
\usepackage{subcaption}

\usepackage{tikz}
\usetikzlibrary{decorations.pathreplacing, calligraphy}
\usetikzlibrary{patterns}
\usetikzlibrary{positioning}
\usepackage{graphicx}

\newcommand{\mcal}{\mathcal}

\renewcommand{\epsilon}{\varepsilon}

\DeclareMathOperator*{\argmax}{\arg\!\max}
\DeclareMathOperator*{\argmin}{\arg\!\min}
\DeclareTextFontCommand{\emph}{\slshape}
\newcommand{\trimax}{\mathop{\max}\limits^{\triangleright}}

\newcommand{\triargmax}{\mathop{\argmax}\limits^{\triangleright}}
\newcommand{\triargmin}{\mathop{\argmin}\limits^{\triangleright}}
\newcommand{\bTheta}{\bm{\Theta}}
\newcommand{\btheta}{\bm{\theta}}

\newtheorem{theorem}{Theorem}
\newtheorem{lemma}{Lemma}
\newtheorem{proposition}{Proposition}

\newtheorem*{statement*}{Result}

\theoremstyle{definition}
\newtheorem{definition}{Definition}

\usepackage{natbib}
\title{Private From Whom? \linebreak A Privacy Frontier for Auctions}

\author{Eric Gao\thanks{Department of Economics, Massachusetts Institute of Technology. ericgao@mit.edu.} 
\space and Eric Tang\thanks{Graduate School of Business, Stanford University. ertang@stanford.edu.} 
\footnote{We thank Itai Ashlagi, Ian Ball, Modibo Camara, Roberto Corrao, Laura Doval, Glenn Ellison, Drew Fudenberg, Daniel Luo, Matthew Gentzkow, Andy Haupt, Jack Hirsch, Zoe Hitzig, Ravi Jagadeesan, Andrew Koh, Andrew Komo, Paul Milgrom, Ellen Muir, Michael Ostrovsky, Parag Pathak, Al Roth, Ilya Segal, Andy Skrzypacz, Takuo Sugaya, Alex Wolitzky, Weijie Zhong, and seminar participants at MIT, Stanford, and Stony Brook for useful conversations and inspiration. \url{Refine.ink} was used to proofread the paper for consistency and clarity. Tang thanks the NSF Graduate Research Fellowship for financial support. This paper was previously circulated under the title ``Private From Whom? Minimal Information Leakage in Auctions.'' Any errors are our own.}}

\date{\today}

\begin{document}
	
\maketitle
	
\begin{abstract}

In many auctions, bidders may value keeping their information private from the auctioneer and other bidders. Among deterministic bilateral communication protocols, revealing less information to bidders requires revealing more information to the auctioneer, and vice versa. A protocol implementing a choice rule is on the \emph{Privacy Frontier} if no alternative protocol reveals less to both bidders and the auctioneer. The sealed-bid first-price auction, the descending auction and the ascending auction are on the Privacy Frontier, but the sealed-bid second-price auction is not. We provide sufficient conditions for protocols to be on the Privacy Frontier and devise alternative frontier protocols.

\end{abstract}

\noindent \textbf{Keywords}:  Extensive-Form Mechanism Design, Privacy, Auctions \\
\noindent \textbf{JEL Codes}: D44, D47, D82.
	
\newpage
\onehalfspacing

\section{Introduction}

Bidders participating in auctions may value keeping their information private from both the auctioneer and other bidders: Their bids reveal their technology level to rivals or reduce their leverage in post-auction negotiations.\footnote{As \cite{rothkopf1990vickrey} describe, ``People of our acquaintance with experience in conducting business are reluctant to reveal their true costs or valuations. ... It could reveal to potential competitors the extent to which a firm's technology was superior. Most important, it could reveal to others with whom the firm must subsequently negotiate precisely how much it can yield. ... Successful oil lease bidders must deal with drilling contractors, rig owners, and so forth. Successful coal lease bidders must deal with equipment suppliers, railroads, and coal purchasers. Successful construction contract bidders must deal with subcontractors and labor unions.'' In the context of spectrum auctions, \cite{milgrom2020clock} write, ``The practical significance of UWP [Unconditional Winner Privacy] is that it may alleviate winners’ concerns about misuse of the revealed information, encourage participation by bidders who find it costly to figure out their exact bid, and allow the auctioneer to conceal winners’ politically sensitive windfalls.'' \cite{bergemann2018transparent} directly studies a model of repeated auctions and finds that minimal disclosure leads to an efficient equilibrium while additional disclosure leads to the existence of an inefficient, low-revenue equilibrium.} Protecting bidders' private information may encourage bidder participation and prevent the auctioneer from learning sensitive information.\footnote{\cite{mcmillan1994spectrum} describes a New Zealand spectrum auction in which the auction revealed a large gap between the winner's willingness to pay and their actual payment, resulting in public criticism of the government. A government auctioneer may wish to avoid criticism by reducing the amount they learn about bidders' values.} We study the following fundamental question: To what extent can different auction formats protect bidders' private information from both the auctioneer and other bidders? 

Consider a designer choosing between a sealed-bid first-price auction and a descending auction. In a sealed-bid first-price auction, the auctioneer observes every bidder’s exact bid, including losing bids. Alternatively, a descending auction seems to better protect bidders' privacy: The auction ends when the highest bidder claims the item, so the auctioneer no longer observes exact losing bids. However, a descending auction reveals more information \textit{to bidders}: Each losing bidder now infers that the winning bid is exactly the price at which the auction stops. In contrast, in a sealed-bid first-price auction bidders only learn whether they won or lost, so losing bidders learn only that the winning bidder's value is higher and the winning bidder learns only that losing values are lower.

What drives this tradeoff? Suppose an auctioneer is running a descending auction and a bidder claims the object. The outcome is already determined, so further questions from the auctioneer are unnecessary yet give the auctioneer more information. But if the auctioneer \textit{stops} asking questions to some bidder, that bidder infers that a competitor won the auction at the last price they were offered. In general, protecting bidders' privacy from the auctioneer requires the auctioneer to not ask bidders questions about their type when the outcome is identical regardless of their answer. However, whether a  question is necessary depends on what the auctioneer already knows about other bidders' types: By conditioning which questions to ask on their current information, the auctioneer reveals other bidders' private information. 

This stark tradeoff holds more generally.
Consider a designer implementing a given choice rule mapping some type space to some set of possible outcomes.\footnote{In the single-unit auction setting, a choice rule describes the allocation of the good and agents' transfers as a function of bidders' types. The first-price auction choice rule and second-price auction choice rule are two examples. We focus on the single-unit auction setting, but our model and main theorems apply to arbitrary choice rules.} To implement the choice rule, the designer chooses a deterministic bilateral communication protocol among an auctioneer and bidders.
Under a general condition on the choice rule, implementation necessitates that for every bidder, (1) the auctioneer learns information about that bidder not pivotal to the outcome or (2) that bidder learns about other bidders' types beyond what the outcome alone reveals.
Consequently, the designer must choose how information leakages are distributed between the auctioneer and bidders. We analyze this understudied additional dimension of privacy---bidders learning about other bidders' private information.

Given this tradeoff, some privacy violations are inevitable. However, some protocols can still be improved upon from a privacy perspective. Consider now a designer choosing between a sealed-bid second-price auction and an ascending auction. In a sealed-bid second-price auction, the auctioneer learns all bidders' values; losing bidders learn that the winning bidder's value is higher than their own; and the winning bidder learns the second-highest value via the price they pay. In an ascending auction, the auctioneer no longer learns the winning bidder's valuation (the auction ends when the second-to-last bidder drops out). However, what bidders learn remains unchanged: Losing bidders still only learn that the winning bidder's value is higher than their own and the winning bidder still learns the second-highest value via the price they pay. The ascending auction \textit{privacy-dominates} the sealed-bid second-price auction: It reveals strictly less information to the auctioneer and the same information to bidders. In general, a protocol lies on the \textit{Privacy Frontier} if it is not privacy-dominated by any protocol implementing the same underlying choice rule.

Given these inevitable tradeoffs, when is a protocol on the Privacy Frontier? Theorem \ref{thm:suff-conditions-frontier}  provides sufficient conditions to determine whether a given protocol lies on the Privacy Frontier. These conditions allow a designer to evaluate a protocol only using model primitives and a description of what information the protocol reveals. Our sufficient conditions apply to any type space and choice rule, enabling analysis beyond the structure of auction settings. 

We apply our framework to study the implementation of first-price auctions and second-price auctions. For the first-price auction choice rule, both the sealed-bid first-price auction and the descending auction lie on the Privacy Frontier.
We then propose two new protocols on the Privacy Frontier. The \textit{descending then sealed-bid} protocol trades off privacy from bidders versus privacy from the auctioneer at different ranges of bidder values. The \textit{descending-holdout} protocol protects one selected bidder's private information from both bidders and the auctioneer.

To implement the second-price auction choice rule, the ascending-join protocol, proposed by \cite{haupt2025contextually}, reveals minimal information to the auctioneer.\footnote{Intuitively, the ascending-join protocol orders the bidders and runs an ascending auction with two bidders at a time, until one bidder drops out, then brings the next bidder into the ascending auction. This prevents the auctioneer from learning the value of later bidders, if later bidders have low values and would lose the auction. We elaborate on this protocol in \cref{sec:second-price-auctions}.} However, other protocols reveal less information to bidders. In particular, the ascending-join protocol does not privacy-dominate the standard ascending protocol, which reveals minimal information to bidders. We show the ascending protocol is on the Privacy Frontier, but the sealed-bid second-price auction is not.

Our results give a designer tools to balance revealing information to the auctioneer with revealing information to bidders. The designer's preferred choice may depend on the specific interaction taking place, which we do not take a stand on. In some settings, such as a one-time auction for a procurement contract, preventing bidders from learning about rival bidders' technological capabilities may be more important. In other settings, such as repeated auctions for a good\footnote{These settings could include auctions for online ads or natural resources.} (in which the auctioneer may set future reserve prices based on bidders' values), preserving privacy from the auctioneer may be more important.

Our framework of deterministic bilateral communication restricts the usage of alternative tools such as cryptography or an external mediator.\footnote{Formally, our designer does not rely on a trusted randomization device as in \cite{izmalkov2011perfect}, and we make no computational hardness assumptions.} These alternative tools may be costly to implement and come with their own limitations. Participants must understand or trust that a cryptographic scheme works as intended for the mechanism to be successful. Relying on an external mediator requires bidders to trust that the mediator will not divulge any information. Finally, alternative tools may be inaccessible, become obsolete, or be prohibitively costly for alternative reasons.\footnote{In many auctions in practice, the auctioneer and bidders communicate directly without a mediator. These include art and antique auctions (at Sotheby's or Christie's), and some auctions for resources such as timber and natural gas. Developments in quantum computing (\cite{cherney2025quantum}) may undermine the security guarantees of cryptographic schemes. In spectrum auctions, where a graph coloring problem must be solved quickly in every iteration, any increase in time complexity could be prohibitive.} With that in mind, our results provide a calculus for deciding \textit{when} the costs of moving to an alternative scheme are worthwhile. A designer may first wish to understand what privacy protections are feasible in the absence of cryptography before deciding whether to use cryptographic protocols; our work provides this characterization.

Our conceptual contribution is to analyze agents' privacy from \emph{other agents}, while most prior literature focuses on privacy from the auctioneer or from an outside observer. Metrics such as contextual privacy (\cite{haupt2025contextually}), relative informativeness (\cite{segal2007communication}, \cite{mackenzie2022menu}), and Bayesian privacy (\cite{eliaz2019constrained}, \cite{eilat2021bayesian}) study the information that a mechanism reveals to the auctioneer or designer. Differential privacy (\cite{dwork2006differential}, \cite{pai2013privacy}) describes the information revealed to an outside observer. Our work instead analyzes the information that a mechanism reveals to \emph{agents} about other agents, and the tradeoff with the information revealed to the auctioneer. This focus is motivated by settings in which privacy violations can affect downstream interactions, such as the models of \cite{dworczak2020aftermarkets} and \cite{rothkopf1990vickrey}, but we provide a privacy metric agnostic to the exact downstream model. \cite{chor1991boolean} and \cite{brandt2008privacy} provide a demanding classification of a mechanism as ``unconditionally private'' if it reveals no unnecessary information about agents, but we describe a Privacy Frontier in the many settings when no unconditionally private mechanism exists.

The remainder of the paper proceeds as follows. \cref{sec:model} introduces our model and defines the Privacy Frontier while \cref{section:FPA_example} works through our notions in a first-price auction setting. \cref{sec:impossibility-theorem} states our impossibility theorem, identifying the tradeoff between revealing information to bidders and the auctioneer. \cref{sec:sufficient-conditions-privacy-frontier} provides sufficient conditions for a protocol to lie on the Privacy Frontier. \cref{sec:first-price-auction} and \cref{sec:second-price-auctions} apply our results to first and second-price auctions respectively. \cref{sec:related-literature} reviews related literature. \cref{sec:conclusion} concludes. All omitted proofs are in \ref{Appendix C}.

\section{Model}\label{sec:model}

Consider a finite set of agents (bidders) $N = \{1, 2, ..., n\}$, with generic element $i$. Each bidder $i$ has value $\theta_i$ in a finite type space $\Theta_i$, drawn independently. $\bm{\Theta} = \times_{i \in N} \Theta_i$ denotes the set of all possible type vectors. $X$ denotes the set of possible outcomes.

A designer, external to the interaction between bidders and the auctioneer, is tasked with implementing an exogenously chosen choice rule $\phi: \bm{\Theta} \to X$.\footnote{The choice of $\phi$ may be motivated by efficiency, revenue, incentive properties of $\phi$, or institutional norms. We do not take a stand on how $\phi$ is determined; we instead consider the tradeoffs faced by a designer choosing how to implement a fixed $\phi$.} In the auction setting, this rule describes agents' transfers and allocation as a function of the type profile. 

For a given choice rule $\phi$, the designer still has considerable flexibility to choose an extensive-form implementation of $\phi$; privacy violations stem from the chosen extensive-form implementation. Formally, the designer chooses an extensive-form game 
$$G = (H, i, \{\mcal I_i\}_{i \in N}, \{A_i\}_{i \in N}, A, I, g)$$
of finite length and perfect recall with notation specified in \cref{tab:notation-extensive-form}.\footnote{We model the designer as external to the interaction between bidders and the auctioneer. In many settings, the designer that advises organizations on what mechanism to use is not a participant themselves, and simply seeks to identify tradeoffs, bounds on what is inherently feasible, and how to best achieve certain objectives.}

\begin{table}[h]
\centering
\caption{Notation for Extensive-Form Games}
\label{tab:notation-extensive-form}
\begin{tabular}{l@{\hskip 0.5cm}l>{\centering\arraybackslash}m{3.5cm}}
\toprule
\textbf{Name} & \textbf{Notation} & \textbf{Representative Element} \\
\midrule
Histories & $H$ & $h$ \\
Player called to play at $h$ & $i(h)$ & \\
Information sets of player $i$ & $\mathcal{I}_i$ & $I_i$ \\
Actions available to $i$ & $A_i$ & \\
Actions available at $I_i$ & $A(I_i)$ & $a$ \\
Information set at history $h$ & $I(h)$ & \\
Terminal histories & $Z$ & $z$ \\
Outcome at terminal node & $g(z)$ & \\
Interim strategy of player $i$ & $\sigma_i(\mathcal{I}_i)$ & \\
\bottomrule
\end{tabular}
\end{table}

We limit our attention to deterministic strategy profiles and abstract away from incentives.\footnote{\ref{Appendix A} discusses pruning protocols and establishes that every protocol implementing a dominant strategy incentive compatible choice rule is ex-post incentive compatible when pruned.} An \emph{interim strategy} for bidder $i$ is a map $\sigma_i:\mathcal{I}_i \to A_i$ satisfying $\sigma_i(I_i) \in A(I_i)$ for all $I_i$. Let $\Sigma_i$ denote the space of interim strategies for bidder $i$. An \emph{ex-ante strategy} for bidder $i$ is a map $S_i: \Theta_i \to \Sigma_i$ from types to interim strategies. A \textit{protocol} $P$ is a pair $P = (G,\bm{S})$ where $\bm{S} = \{S_i\}_{i \in N}$ is a fixed profile of ex-ante strategies for $G$.

Given $S$, each type profile $\bm{\theta}$ yields a sequence of information sets and bidder actions at those information sets. Let $I_i^k(\bm \theta)$ denote the $k$th information set reached at which bidder $i$ is called to play, and let $a_i^k(\bm \theta)$ denote their action chosen at this information set. Let $I^k(\bm \theta)$ denote the $k$th information set reached in the game, and let $a^k(\bm \theta)$ denote the action chosen at this information set. Let $z(\{\sigma_i\}_{i \in N})$ denote the terminal node reached via playing interim strategies $\{\sigma_i\}_{i \in N}$ so $z(\bm S(\bm\theta))$ is the terminal node reached via playing ex-ante strategies $\bm S$ at type profile $\bm \theta$. Protocol $P$ implements social choice rule $\phi$ if for all $\bm \theta$, $\phi(\bm \theta) = g(z(\bm S(\bm \theta))).$

At this stage, we do not restrict the incentive properties of $P$. This only strengthens our two main results. \cref{thm:impossibility-thm}, our impossibility theorem, describes when any protocol $P$ implementing $\phi$ yields privacy violations -- this holds whether or not $P$ satisfies a given equilibrium concept. \cref{thm:suff-conditions-frontier} provides conditions for a protocol to lie on the Privacy Frontier. Protocol $P$ lies on the Privacy Frontier if it is not privacy-dominated by any protocol implementing $\phi$; such a $P$ therefore is not dominated by any protocol satisfying given incentive properties. In \cref{sec:first-price-auction} and \cref{sec:second-price-auctions} we apply our main results and provide concrete protocols that lie on the Privacy Frontier. We separately check that these protocols also satisfy desirable incentive properties.\footnote{In the second-price auction setting, these protocols are ex-post incentive compatible or dominant-strategy incentive compatible. In the first-price auction setting, these protocols form a perfect Bayesian equilibrium given adequate utility and prior-dependent bidding functions.} 

What does the auctioneer observe from the protocol? The auctioneer implements the choice rule by repeatedly querying a bidder, observing their action, and moving to the corresponding successor information set; this continues until the auctioneer arrives at a terminal history. This embeds an assumption that the auctioneer cannot ``commit to forget'' the content of a bidder's message. Hence the auctioneer observes the full sequence of information sets reached and actions taken during the play of the game: For a given protocol $P$ and type profile $\bm{\theta}$, the auctioneer observes the game history $(I^k(\bm \theta),a^k(\bm \theta))_{k=1}^{T}$. Let $o_A(P,\bm{\theta})$ denote this observed history under protocol $P$ at type profile $\theta$. We say $\bm{\theta} \sim_{(P, A)} \bm{\theta}'$ if $o_A(P, \bm{\theta}) = o_A(P, \bm{\theta}')$: If the auctioneer cannot distinguish between type profiles $\bm{\theta}$ and $\bm{\theta}'$, then they are equivalent under $\sim_{(P, A)}$. To implement $\phi$, the auctioneer must \textit{at least} be able to distinguish between type profiles that prescribe different outcomes. Letting $\mcal P(\sim)$ denote the partition induced by an equivalence relation $\sim$, this condition can be stated in the language of partitions: $\mcal P(\sim_{(P, A)})$ must be finer than $\mcal P(\sim_\phi)$ where $a \sim_\phi b$ if and only if $\phi(a) = \phi(b)$.\footnote{Elements $a, b \in \Theta$ are defined to be in the same cell of $\mcal P(\sim)$ if and only if $a \sim b$. Partition $\mathcal{P}$ is finer than partition $\mathcal{P}'$ if for any cell $X$ in $\mcal P$, there exists a cell $Y$ in $\mcal P'$ such that $X \subseteq Y$.} However, if $\mcal P(\sim_{(P, A)})$ is strictly finer than $\mcal P(\sim_\phi)$, the auctioneer learns more than what is needed to implement $\phi$.\footnote{This is a ``contextual privacy violation'' in the language of \cite{haupt2025contextually}. We choose our terminology to distinguish these privacy violations from bidders learning unnecessary information about one another.} If this is the case, $P$ \textit{reveals unnecessary information to the auctioneer}.\footnote{In addition, $P$ reveals unnecessary information to the auctioneer \emph{about bidder $i$} if there exists $\theta_i, \theta_i' \in \Theta_i$ and $\btheta_{-i} \in \bTheta_{-i}$ with $\phi(\theta_i, \btheta_{-i}) = \phi(\theta_i', \btheta_{-i})$ but $(\theta_i, \btheta_{-i}) \nsim_{(P, A)} (\theta_i', \btheta_{-i})$.}

What do bidders observe from the protocol? First, the bidder can distinguish some aspect of the outcome and thus make inferences about others' types based on that. Let $\Omega_i$ be a partition of $X$, with cells denoted by $\omega_i$, describing the outcomes that bidder $i$ can distinguish from one another. In the auction setting, we suppose that a bidder observes their allocation and payment at the conclusion of the auction.\footnote{Suppose $\{(q_j,t_j)\}_{j \in N} \in X$ represents an outcome with allocation $q_i \in \{0,1\}$ and transfer $t_i$ for bidder $i$. Then $\{(q_j,t_j)\}_{j \in N} \sim_{\Omega_i} \{(q_j',t_j')\}_{j=1}^N$ if $q_i = q_i'$ and $t_i = t_i'$. This corresponds to bidder $i$ being unable to inherently distinguish between two different outcomes which both result in them losing the auction.} Furthermore, each bidder observes their own sequence of communication with the auctioneer. In total, bidder $i$ observes $(\theta_i, (I_i^k(\bm \theta),a_i^k(\bm \theta))_{k=1}^{T_i}, \omega_i)$: the sequence of the information sets at which they have been called to play, the actions they have taken at those information sets, and what they observe about the final outcome, $\omega_i\in\Omega_i$.\footnote{Under this formulation, the bidder can observe their sequence of play, but not calendar time. This is also the definition of an experience, restricted to the end of the game, in \cite{li2025obvious}.} Let $o_i(P,\bm{\theta})$ denote bidder $i$'s observation in protocol $P$ when bidders have type profile $\bm{\theta}\in\bm{\Theta}$. If $o_i(P, \bm{\theta}) = o_i(P, \bm{\theta}')$ then $\bm{\theta} \sim_{(P,i)} \bm{\theta}'$ and $\bm{\theta}$ is observationally equivalent to $\bm{\theta}'$. Similarly, $\sim_{(P, i)}$ induces a partition $\mcal P(\sim_{(P, i)})$ over $\bm \Theta$ for bidder $i$. If $\mcal P(\sim_{(P, i)})$ is strictly finer than what bidder $i$ learns about other bidders based on observing the outcome of the auction, the protocol \textit{reveals unnecessary information to bidder $i$}. This happens if there exists $\theta_i, \bm{\theta}_{-i}, \bm{\theta}_{-i}'$ such that $\phi(\theta_i, \bm{\theta}_{-i}) \sim_{\Omega_i} \phi(\theta_i, \bm{\theta}_{-i}')$ but $(\theta_i, \bm{\theta}_{-i}) \not \sim_{(P, i)} (\theta_i, \bm{\theta}_{-i}')$. Unnecessary revelations of information constitute our notion of privacy violations.

\subsection{Example: First-Price Auction}\label{section:FPA_example}

We illustrate our definitions of the information revealed to the auctioneer and to bidders with a simple first-price auction.  \cref{fig:fpa-2v2} represents a first-price auction with two bidders and two types, with values $\Theta_1 = \Theta_2 = \{0,1\}$ and discrete bid space $\{0, \frac{1}{2}\}$, in which bidder $1$ wins ties.\footnote{When values are distributed uniformly on $\Theta$, the  bids $b_i(0) = 0$ and $b_i(1) = \frac{1}{2}$ are incentive-compatible.} The parenthetical terms in each cell represent the outcomes $(i, p)$ of the choice rule $\phi^{\text{FPA}}$, with $i$ being the winner of the auction and $p$ the price paid by the winner. The possible outcomes are $X = \{(1,\frac{1}{2}), (1,0), (2, \frac{1}{2})\}$. Recall that agent $i$ can distinguish two outcomes if they result in different allocations or transfers for agent $i$. Agent $2$ cannot distinguish the price that agent $1$ pays: $(1, 0) \sim_{\Omega_2} (1,\frac{1}{2})$. For all other $x, x' \in X$ and $i \in N$, $x \nsim_{\Omega_i} x'$. In \cref{fig:fpa-2v2}, distinct colors represent distinct outcomes in $X$, and distinct shades of blue represent outcomes that agent $2$ cannot distinguish.

\begin{figure}[h]
\centering
\begin{tikzpicture}[scale=0.8]

\fill[blue, opacity=0.2] (0,0) rectangle (3,3);
\fill[blue, opacity=0.4] (3,0) rectangle (6,3);
\fill[red, opacity=0.4] (0,3) rectangle (3,6);
\fill[blue, opacity=0.4] (3,3) rectangle (6,6);

\draw[thick] (0,0) rectangle (6,6);

\draw[dashed] (3,0) -- (3,6);
\draw[dashed] (0,3) -- (6,3);

\node at (1.5,1.5) {$(1,0)$};
\node at (4.5,1.5) {$(1,\frac{1}{2})$};
\node at (1.5,4.5) {$(2,\frac{1}{2})$};
\node at (4.5,4.5) {$(1,\frac{1}{2})$};

\node[below] at (1.5,0) {$\theta_1=0$};
\node[below] at (4.5,0) {$\theta_1=1$};

\node[left] at (0,1.5) {$\theta_2 = 0$};
\node[left] at (0,4.5) {$\theta_2 = 1$};

\end{tikzpicture}
\caption{First-Price Auction with 2 Bidders}
\label{fig:fpa-2v2}
\end{figure}

Two natural protocols to implement $\phi^{\text{FPA}}$ are a sealed-bid protocol and a descending protocol. Under a descending protocol, denoted $P^D$, the auctioneer: 
\begin{enumerate}
    \item Asks agent one if $\theta_1 = 1$. If yes, output $(1, \frac{1}{2})$. Else:
    \item Asks agent two if $\theta_2 = 1$. If yes, output $(2, \frac{1}{2})$. Else: 
    \item Outputs $(1, 0)$.
\end{enumerate}

One equilibrium is for agents to truthfully answer the auctioneer's question at each information set. In the descending protocol, the auctioneer always learns the type of agent $1$, but only learns the type of agent $2$ if $\theta_1 = 0$. This yields the partition $\mathcal{P}(\sim_{(P^D, A)})$ displayed in \cref{fig:information-revealed-PD}. Now consider $\mathcal{P}(\sim_{(P^D, 2)})$, the information revealed to agent $2$. Agent $2$ always observes their own type. They also distinguish type profile $(1, 1)$ from $(0, 1)$, as they only win in the latter type profile. Finally, at type profile $(1,0)$, agent 2 never arrives at an information set while at type profile $(0,0)$, they arrive at an information set, so $(1,0) \nsim_{(P^D, 2)} (0,0)$. Hence the mechanism reveals the exact type profile to agent $2$, as in \cref{fig:information-revealed-PD}. If $\theta_2 = 0$, agent two does not learn agent one's type from the outcome so bidder $2$ learns unnecessary information about other bidders under a descending protocol.

\begin{figure}[h]
\centering

\begin{subfigure}[b]{0.45\textwidth}
    \centering
    \begin{tikzpicture}[scale=0.8]
    \fill[blue, opacity=0.2] (0,0) rectangle (3,3);
    \fill[blue, opacity=0.4] (3,0) rectangle (6,3);
    \fill[red, opacity=0.4] (0,3) rectangle (3,6);
    \fill[blue, opacity=0.4] (3,3) rectangle (6,6);
    \draw[thick] (0,0) rectangle (6,6);
    \draw[ultra thick] (3,0) -- (3,6);
    \draw[ultra thick] (0,3) -- (3,3);
    \node at (1.5,1.5) {$(1,0)$};
    \node at (4.5,1.5) {$(1,\frac{1}{2})$};
    \node at (1.5,4.5) {$(2,\frac{1}{2})$};
    \node at (4.5,4.5) {$(1,\frac{1}{2})$};
    \node[below] at (1.5,0) {$\theta_1=0$};
    \node[below] at (4.5,0) {$\theta_1=1$};
    \node[left] at (0,1.5) {$\theta_2 = 0$};
    \node[left] at (0,4.5) {$\theta_2 = 1$};
    \end{tikzpicture}
    \caption{$\mathcal{P}(\sim_{(P^D, A)})$}
\end{subfigure}%
\hfill
\begin{subfigure}[b]{0.45\textwidth}
    \centering
    \begin{tikzpicture}[scale=0.8]
    \fill[blue, opacity=0.2] (0,0) rectangle (3,3);
    \fill[blue, opacity=0.4] (3,0) rectangle (6,3);
    \fill[red, opacity=0.4] (0,3) rectangle (3,6);
    \fill[blue, opacity=0.4] (3,3) rectangle (6,6);
    \draw[thick] (0,0) rectangle (6,6);
    \draw[ultra thick] (3,0) -- (3,6);
    \draw[ultra thick] (0,3) -- (6,3);
    \node at (1.5,1.5) {$(1,0)$};
    \node at (4.5,1.5) {$(1,\frac{1}{2})$};
    \node at (1.5,4.5) {$(2,\frac{1}{2})$};
    \node at (4.5,4.5) {$(1,\frac{1}{2})$};
    \node[below] at (1.5,0) {$\theta_1=0$};
    \node[below] at (4.5,0) {$\theta_1=1$};
    \node[left] at (0,1.5) {$\theta_2 = 0$};
    \node[left] at (0,4.5) {$\theta_2 = 1$};
    \end{tikzpicture}
    \caption{$\mathcal{P}(\sim_{(P^D, 2)})$}
\end{subfigure}
\caption{Information Revealed to Auctioneer and Bidder 2 by $P^D$}
\label{fig:information-revealed-PD}
\end{figure}

Alternatively, the designer could use a sealed-bid first-price auction, in which each agent reveals their type to the auctioneer, and the auctioneer then chooses the outcome. This yields the partitions in \cref{fig:information-revealed-sealed-bid-fpa}. The auctioneer learns the exact type profile $\bm{\theta}$. Since the auctioneer does not need to distinguish type profiles $(1, 1)$ from $(1, 0)$, the auctioneer learns unnecessary information under a sealed bid protocol. Agent $2$ observes whether they win the auction, so if $\theta_2=1$, they distinguish the type $\theta_1$. However, if $\theta_2=0$, all agent $2$ observes is that they lost the auction, hence they cannot distinguish the type $\theta_1$.

\begin{figure}[h]
\centering

\begin{subfigure}[b]{0.45\textwidth}
    \centering
    \begin{tikzpicture}[scale=0.8]
    \fill[blue, opacity=0.2] (0,0) rectangle (3,3);
    \fill[blue, opacity=0.4] (3,0) rectangle (6,3);
    \fill[red, opacity=0.4] (0,3) rectangle (3,6);
    \fill[blue, opacity=0.4] (3,3) rectangle (6,6);
    \draw[thick] (0,0) rectangle (6,6);
    \draw[ultra thick] (3,0) -- (3,6);
    \draw[ultra thick] (0,3) -- (6,3);
    \node at (1.5,1.5) {$(1,0)$};
    \node at (4.5,1.5) {$(1,\frac{1}{2})$};
    \node at (1.5,4.5) {$(2,\frac{1}{2})$};
    \node at (4.5,4.5) {$(1,\frac{1}{2})$};
    \node[below] at (1.5,0) {$\theta_1=0$};
    \node[below] at (4.5,0) {$\theta_1=1$};
    \node[left] at (0,1.5) {$\theta_2 = 0$};
    \node[left] at (0,4.5) {$\theta_2 = 1$};
    \end{tikzpicture}
    \caption{$\mathcal{P}(\sim_{(P^{\text{SB}}, A)})$}
\end{subfigure}%
\hfill
\begin{subfigure}[b]{0.45\textwidth}
    \centering
    \begin{tikzpicture}[scale=0.8]
    \fill[blue, opacity=0.2] (0,0) rectangle (3,3);
    \fill[blue, opacity=0.4] (3,0) rectangle (6,3);
    \fill[red, opacity=0.4] (0,3) rectangle (3,6);
    \fill[blue, opacity=0.4] (3,3) rectangle (6,6);
    \draw[thick] (0,0) rectangle (6,6);
    \draw[ultra thick] (3,3) -- (3,6);
    \draw[ultra thick] (0,3) -- (6,3);
    \node at (1.5,1.5) {$(1,0)$};
    \node at (4.5,1.5) {$(1,\frac{1}{2})$};
    \node at (1.5,4.5) {$(2,\frac{1}{2})$};
    \node at (4.5,4.5) {$(1,\frac{1}{2})$};
    \node[below] at (1.5,0) {$\theta_1=0$};
    \node[below] at (4.5,0) {$\theta_1=1$};
    \node[left] at (0,1.5) {$\theta_2 = 0$};
    \node[left] at (0,4.5) {$\theta_2 = 1$};
    \end{tikzpicture}
    \caption{$\mathcal{P}(\sim_{(P^{\text{SB}}, 2)})$}
\end{subfigure}
\caption{Information Revealed to Auctioneer and Bidder 2 by $P^{\text{SB}}$}
\label{fig:information-revealed-sealed-bid-fpa}
\end{figure}

As noted above, these bidding functions are incentive compatible when, for example, values are distributed uniformly. When analyzing incentive compatibility here and in our applications, we do not model how bidders' privacy concerns affect their incentives in the mechanism. For example, when considering a sealed-bid second-price auction in \cref{sec:second-price-auctions}, we analyze the protocol in which each bidder simply bids his value. One motivation is that our analysis is agnostic to the nature of downstream interactions between agents: distinct downstream models could lead agents to distort their behavior in very distinct ways. We do not take a stand on the nature of the downstream interaction, or how agents trade off privacy with other objectives, and instead propose an agnostic definition.\footnote{A separate motivation for this assumption may be a ``paternalistic'' model of privacy, as in \cite{eliaz2019constrained} and \cite{eilat2021bayesian}, in which consumers are not fully aware of the effects of privacy loss.}

\subsection{The Privacy Frontier}

As we show later in \cref{thm:impossibility-thm}, typically no one protocol is best at protecting privacy from both the auctioneer and from bidders. However, protocols can still be partially ordered by the information they reveal to the auctioneer and to bidders. Protocol $P$ reveals more to bidder $i$ than protocol $P'$ if bidder $i$ can distinguish any pairs of type profiles in $P$ that they could distinguish in $P'$. A similar definition holds for what the auctioneer learns.

\begin{definition}
    Protocol $P$ \textit{reveals (strictly) more to bidder} $i$ than protocol $P'$ if $\mcal P(\sim_{(P, i)})$ is (strictly) finer than $\mcal P(\sim_{(P', i)})$. Protocol $P$ \textit{reveals (strictly) more to the auctioneer} than protocol $P'$ if $\mcal P(\sim_{(P, A)})$ is (strictly) finer than $\mcal P(\sim_{(P', A)})$.
\end{definition}

One protocol privacy-dominates another if it reveals less information to the auctioneer and to bidders. If a protocol is not dominated by any other protocol that implements the same choice rule, it lies on the Privacy Frontier. 

\begin{definition}
    Protocol $P'$ \emph{privacy-dominates} protocol $P$ if $P$ reveals more to bidder $i$ than $P'$ for all $i\in N$, and $P$ reveals more to the auctioneer than $P'$. $P'$ \emph{strictly privacy-dominates} $P$ if $P'$ privacy-dominates $P$ and reveals strictly less information than $P$ to the auctioneer or to some bidder $i \in N$.
\end{definition}

\begin{definition}
    A protocol $P$ implementing $\phi$ lies on the \emph{Privacy Frontier} if there does not exist a protocol implementing $\phi$ which strictly privacy-dominates $P$.
\end{definition}

Since partitions are only partially ordered, ordering protocols by bidder or auctioneer privacy is also a partial ordering. As such, the Privacy Frontier will generally have many incomparable protocols on the frontier. For example, consider the first-price auction $\phi^{\text{FPA}}$ and the two implementations, $P^{\text{D}}$ and $P^{\text{SB}}$, described in \cref{fig:fpa-2v2} through \cref{fig:information-revealed-sealed-bid-fpa}. Neither $P^{\text{D}}$ nor $P^{\text{SB}}$ privacy-dominates the other protocol. This is because $P^{\text{D}}$ reveals strictly less to the auctioneer but $P^{\text{SB}}$ reveals strictly less to agent $2$. We now show this privacy tradeoff is more general than in these two protocols, and provide a systematic analysis of the Privacy Frontier.

\section{Privacy}\label{sec:privacy}

\subsection{Unavoidable Privacy Tradeoffs}\label{sec:impossibility-theorem}

To implement many choice rules, a designer must trade off between protecting privacy from the auctioneer and privacy from bidders. \cref{thm:impossibility-thm} shows that revealing less information to the auctioneer often conflicts with revealing less information to agents. Formally: 

\begin{theorem}\label{thm:impossibility-thm}
    For any choice rule $\phi$, types $\theta_i, \theta_i' \in \Theta_i$ and $\bm{\theta}_{-i}, \bm{\theta}_{-i}' \in \bm{\Theta}_{-i}$ with $\phi(\theta_i, \bm{\theta}_{-i}') \neq \phi(\theta_i', \bm{\theta}_{-i}')$, no protocol $P$ implementing $\phi$ can satisfy both $(\theta_i, \bm{\theta}_{-i}) \sim_{(P, A)} (\theta_i', \bm{\theta}_{-i})$ and
    $(\theta_i, \bm{\theta}_{-i}) \sim_{(P, i)} (\theta_i, \bm{\theta}_{-i}')$. Thus, if additionally $\phi(\theta_i, \bm{\theta}_{-i}) = \phi(\theta_i', \bm{\theta}_{-i})$ and $\phi (\theta_i, \bm{\theta}_{-i}) \sim_{\Omega_i} \phi(\theta_i, \bm{\theta}_{-i}')$, then any protocol implementing $\phi$ must reveal unnecessary information \textit{to the auctioneer} about bidder $i$ or reveal unnecessary information \textit{to bidder $i$}.
\end{theorem}

If a choice rule $\phi$ satisfies the three conditions of \cref{thm:impossibility-thm} at types $\theta_i, \theta_i'$ and $\bm{\theta}_{-i}, \bm{\theta}_{-i}'$ (i.e. $\phi(\theta_i, \bm{\theta}_{-i}') \neq \phi(\theta_i', \bm{\theta}_{-i}'), \phi(\theta_i, \bm{\theta}_{-i}) = \phi(\theta_i', \bm{\theta}_{-i})$ and $\phi (\theta_i, \bm{\theta}_{-i}) \sim_{\Omega_i} \phi (\theta_i, \bm{\theta}_{-i}')$), we say $\phi$ violates the \textit{indistinguishable corners condition} at types $\theta_i, \theta_i'$ and $\bm{\theta}_{-i}, \bm{\theta}_{-i}'$. This condition generalizes the corners condition introduced by \cite{chor1991boolean} to further account for what bidders learn. As such, \cref{thm:impossibility-thm} tells us that if $\phi$ has any violations of the indistinguishable corners condition, no protocol can simultaneously (1) reveal no unnecessary information to the auctioneer and (2) reveal no unnecessary information to the bidders. As such, simply choosing a protocol which maximally preserves privacy from the auctioneer, such as the descending auction or the ascending-join auction of \cite{haupt2025contextually}, may not be optimal for a privacy-conscious designer. In particular, if a protocol $P$ does not reveal any unnecessary information to the auctioneer about a given bidder, we can produce a protocol that reveals less unnecessary information to bidders than $P$ (such as a sealed-bid protocol). This motivates the Privacy Frontier being nontrivial---no one protocol dominates all other protocols---and complicates the privacy argument for using certain protocols without additionally considering privacy \textit{from whom}.

To illustrate \cref{thm:impossibility-thm}, consider the first-price auction of \cref{fig:fpa-2v2}. Outcome $\phi(0, 0) \neq \phi(0, 1)$, as the type of agent $2$ determines the winner. \cref{thm:impossibility-thm} states that \emph{no} protocol $P$ implementing $\phi^{\text{FPA}}$ can satisfy both $(1, 0) \sim_{(P, A)} (1,1)$ and $(0, 0) \sim_{(P, 2)} (1,0)$. In words, any protocol that implements $\phi^{\text{FPA}}$ must reveal at least one of the following: (a) if agent $1$ has a high value, the auctioneer learns the value of agent $2$, (b) if agent $2$ has a low value, agent $2$ learns the value of agent $1$. The tradeoff illustrated by the sealed-bid auction and descending auction is therefore not limited to these two formats. Privacy from the auctioneer and privacy from agents inevitably conflict.

What drives this tradeoff in \cref{thm:impossibility-thm}? To prevent agents from learning about other agents, the auctioneer must ask unnecessary questions to make an agent's experience identical across different type realizations. Suppose protocol $P = (G, \bm{S})$ implements choice rule $\phi$. Because $\phi(\theta_i, \bm{\theta}_{-i}') \neq \phi(\theta_i', \bm{\theta}_{-i}')$, if the type profile is $(\theta_i, \bm{\theta}_{-i}')$, the auctioneer must eventually ask agent $i$ to distinguish whether their type is $\theta_i$ or $\theta_i'$, leading agent $i$ to eventually face an information set $I$ at which $S_i(\theta_i)(I) \neq S_i(\theta_i')(I)$. Now suppose that $(\theta_i, \bm{\theta}_{-i}) \sim_{(P, A)} (\theta_i', \bm{\theta}_{-i})$. Then the auctioneer does not distinguish $\theta_i$ from $\theta_i'$ at this type profile: that is, at type profile $(\theta_i, \bm{\theta}_{-i})$, at any information set $\tilde{I}$ reached by agent $i$, $S_i(\theta_i)(\tilde{I}) = S_i(\theta_i')(\tilde{I})$. But then $I$ is not on the path of play for $i$ when the type profile is $(\theta_i, \bm{\theta}_{-i})$. Hence agent $i$ has different experiences at the two type profiles, and so  $(\theta_i, \bm{\theta}_{-i}) \nsim_{(P, i)} (\theta_i, \bm{\theta}_{-i}')$. 

Intuitively, if the auctioneer never asks bidder $i$ to distinguish $\theta_i$ and $\theta_i'$, then bidder $i$ understands the type profile is \emph{not} $(\theta_i,\bm{\theta}_{-i}')$. This is additional information that bidder $i$ obtains from the protocol being run. This yields a tradeoff: given type profile $\bm{\theta}_{-i}$, either the auctioneer never distinguishes between $\theta_i$ and $\theta_i'$ (which reveals information to bidder $i$), or the auctioneer eventually distinguishes between $\theta_i$ and $\theta_i'$ (which reveals information to the auctioneer). This logic yields \cref{lem:auctioneer-distinguishes-both-rows}, which delivers \cref{thm:impossibility-thm}.\footnote{We also make use of the contrapositive of \cref{lem:auctioneer-distinguishes-both-rows}: if $(\theta_i,\bm{\theta}_{-i}) \sim_{(P, i)} (\theta_i,\bm{\theta}_{-i}')$, then $(\theta_i,\bm{\theta}_{-i}) \sim_{(P, A)} (\theta_i',\bm{\theta}_{-i})$ if and only if $(\theta_i,\bm{\theta}_{-i}') \sim_{(P, A)} (\theta_i',\bm{\theta}_{-i}')$.}

\begin{lemma}\label{lem:auctioneer-distinguishes-both-rows}
    Consider $\theta_i,\theta_i'\in\Theta_i$ and $\bm{\theta}_{-i},\bm{\theta}_{-i}' \in \bm{\Theta}_{-i}$. If $(\theta_i,\bm{\theta}_{-i}) \sim_{(P, A)} (\theta_i',\bm{\theta}_{-i})$ and \\ $(\theta_i,\bm{\theta}_{-i}') \nsim_{(P, A)} (\theta_i',\bm{\theta}_{-i}')$ then $(\theta_i,\bm{\theta}_{-i}) \nsim_{(P, i)} (\theta_i,\bm{\theta}_{-i}')$ and $(\theta_i',\bm{\theta}_{-i}) \nsim_{(P, i)} (\theta_i',\bm{\theta}_{-i}')$.
\end{lemma}

These results also cast a new light on existing measures of privacy. Indeed, by the standard notions of relative informativeness (\cite{mackenzie2022menu}) or contextual privacy (\cite{haupt2025contextually}), protocol $P^{\text{D}}$ is more private than $P^{\text{SB}}$. Our analysis suggests that a privacy-conscious designer may prefer $P^{\text{SB}}$, as it reveals less information to bidders. Furthermore, \cref{thm:impossibility-thm} shows that this tradeoff is not limited to $P^{\text{D}}$ and $P^{\text{SB}}$: privacy from the auctioneer and from bidders inherently conflict for any potential protocol.

\subsection{Sufficient Conditions for the Privacy Frontier}\label{sec:sufficient-conditions-privacy-frontier}

In light of the tradeoffs between preserving privacy from the auctioneer and privacy from bidders, how can a designer verify that a given protocol lies on the Privacy Frontier? We provide two economically interpretable conditions which, if satisfied, imply that a protocol is on the Privacy Frontier. Additionally, these conditions only depend on model primitives and the information revealed by a candidate protocol. Furthermore, these conditions apply to implementing any choice function, even beyond auctions. The two conditions are as follows. 

\begin{definition}
    A protocol $P$ implementing choice rule $\phi$ satisfies \textbf{Ex-Post Potential Pivotality} if each bidder can never rule out that an action he took was pivotal to determining the outcome. Formally, for all $i \in N$, all $\theta_i, \theta_i' \in \Theta_i$ and all $\bm{\theta}_{-i} \in \bm{\Theta}_{-i}$ with $(\theta_i, \bm{\theta}_{-i}) \nsim_{(P, A)} (\theta_i', \bm{\theta}_{-i})$, there exists $\bm{\theta}_{-i}' \in \bm{\Theta}_{-i}$ such that:
    \begin{enumerate}[(a)]
        \item $i$ cannot rule out type profile $\bm{\theta}_{-i}'$: $(\theta_i, \bm{\theta}_{-i}) \sim_{(P, i)} (\theta_i, \bm{\theta}_{-i}')$ or $(\theta_i', \bm{\theta}_{-i}) \sim_{(P, i)} (\theta_i', \bm{\theta}_{-i}')$,
        \item $i$ is pivotal at $\bm{\theta}_{-i}'$: $\phi(\theta_i, \bm{\theta}_{-i}') \neq \phi(\theta_i', \bm{\theta}_{-i}')$.
    \end{enumerate}
\end{definition}

To interpret this condition, its definition considers any two type profiles, $(\theta_i, \btheta_{-i})$ and $(\theta_i', \btheta_{-i})$, that the auctioneer distinguishes. The condition is satisfied if agent $i$ believes this distinction \emph{may} have been pivotal at $\theta_i$ or $\theta_i'$: at some profile $\bm{\theta}_{-i}'$, agent $i$'s report affects the final outcome, and $i$ cannot rule out that the true  profile was indeed $\bm{\theta}_{-i}'$ (even after the mechanism has concluded). If $P$ satisfies Ex-Post Potential Pivotality then every query either (1) contributes to determining the outcome or (2) plays a role in homogenizing some bidder's experience between two branches of the extensive form.

We now define a condition on how much information a protocol reveals to bidders. For bidder $i$ and type profile $(\theta_i,\bm{\theta}_{-i})\in\bm{\Theta}$, let $\Pi_i^P(\theta_i, \bm{\theta}_{-i}) := \left\{\widehat\theta_i\in\Theta_i: (\theta_i,\bm{\theta}_{-i})\sim_{(P, A)}(\widehat\theta_i,\bm{\theta}_{-i})\right\}$ be the set of agent-$i$ types that the
auctioneer cannot distinguish from $\theta_i$ when other agents' types are
$\bm{\theta}_{-i}$.

\begin{definition}\label{def:no-avoidable-bidder-distinctions}
A protocol $P$ implementing $\phi$ satisfies \textbf{No Avoidable Bidder
Distinctions} if for all $i\in N$, all $\theta_i\in\Theta_i$, and
all $\bm{\theta}_{-i},\bm{\theta}_{-i}'\in\bm{\Theta}_{-i}$, if $\phi(\theta_i,\bm{\theta}_{-i}) \sim_{\Omega_i} \phi(\theta_i,\bm{\theta}_{-i}')$ and $\Pi_i^P(\theta_i, \bm{\theta}_{-i}) =\Pi_i^P(\theta_i, \bm{\theta}_{-i}')$, then $(\theta_i,\bm{\theta}_{-i}) \sim_{(P, i)} (\theta_i,\bm{\theta}_{-i}')$.
\end{definition}

In words, whenever two profiles of other agents' types result in (a) the auctioneer learning identical information about agent $i$ and (b) agent $i$ observing identical information about the outcome, then agent $i$ is unable to distinguish the two type profiles. To isolate information conveyed by the protocol itself,
\cref{def:no-avoidable-bidder-distinctions} first holds fixed bidder $i$'s own type and requires that the two realized outcomes be observationally equivalent for bidder $i$.  It then asks whether the auctioneer distinguishes $\theta_i$ from exactly the same set of alternative types of bidder $i$ at the two others' profiles. If so, the protocol is not permitted to let bidder $i$ distinguish those others' profiles. Hence changes in query wording, query timing, or the sequence of questions cannot be used to transmit information about other bidders, unless those changes accompany a genuine change in what the auctioneer learns about bidder $i$.

\begin{theorem}\label{thm:suff-conditions-frontier}
    If protocol $P$ satisfies Ex-Post Potential Pivotality and No Avoidable Bidder Distinctions then $P$ lies on the Privacy Frontier.
\end{theorem}

\cref{thm:suff-conditions-frontier} is a general tool for verifying if a protocol is on the Privacy Frontier, which we use throughout our applications. It suffices for a designer to check that: 
\begin{enumerate}[i.]
    \item Any information revealed to the auctioneer is supported by some potentially pivotal type profile (to satisfy Ex-Post Potential Pivotality);
    \item Any information revealed to the bidder is supported by the auctioneer learning differentially (to satisfy No Avoidable Bidder Distinctions).
\end{enumerate}

To illustrate this machinery, we show that the sealed-bid first-price auction of Section \ref{section:FPA_example} is on the Privacy Frontier. By Theorem \ref{thm:suff-conditions-frontier}, it suffices to show that the sealed-bid protocol satisfies Ex-Post Potential Pivotality and No Avoidable Bidder Distinctions. 

\emph{Ex-Post Potential Pivotality.} The auctioneer distinguishes $\theta_1$ from $\theta_1'$ and distinguishes $\theta_2$ from $\theta_2'$. In the sealed-bid first-price auction, agents cannot rule out pivotality. For $i=1$, let $\bm{\theta}_{-i}'$ be $0$ or $1$; then agent $1$ is always pivotal. For $i=2$, let $\bm{\theta}_{-i}'$ be $0$. When $\theta_1=0$, agent $2$ is pivotal to the outcome, as $\phi(0,0)$ has agent $1$ win and $\phi(0, 1)$ has agent $2$ win. Further, $(0, 0) \sim_{(P, 2)} (1, 0)$, because agent $2$ loses in both cases and cannot distinguish agent $1$'s value.

\emph{No Avoidable Bidder Distinctions} is automatically satisfied because each bidder has only one information set. As such, they learn about other bidders only via observing the outcome.

While our sufficiency conditions are only on the partitions induced by candidate protocols, the complications of the full extensive-form have bite. As such, Theorem \ref{thm:suff-conditions-frontier} only pins down sufficient conditions for a protocol to be on the frontier. \ref{Appendix B} provides examples showing that neither condition is necessary.

These complications are resolved when we restrict to static mechanisms. In a \emph{static mechanism}, each bidder has exactly one information set, reached by every type realization. Every static mechanism automatically satisfies no avoidable bidder distinctions since bidders make no distinctions from the protocol (no questions are contingent on any other bidders' private information). Furthermore:

\begin{proposition}\label{prop:static-protocol-privacy-frontier-condition}
    A static protocol $P$ is on the Privacy Frontier if and only if $P$ satisfies ex-post potential pivotality.
\end{proposition}

Theorem \ref{thm:suff-conditions-frontier} automatically implies the backwards direction. Conversely, if there are any violations of ex-post potential pivotality, some queries to an agent that are not pivotal can be feasibly removed by querying this agent last. For non-static protocols, it may not be feasible to remove a query in this way while preserving a valid extensive-form game tree.

\section{First-Price Auctions}\label{sec:first-price-auction}

In this section, we consider an auctioneer implementing a first-price auction choice rule. As discussed in the introduction, a sealed-bid first-price auction reveals more information to the auctioneer, but a descending auction reveals more information to bidders. We first show that a first-price auction choice rule violates the indistinguishable corners condition for every bidder, so the impossibility result of \cref{thm:impossibility-thm} applies. We then show that the descending auction and sealed-bid first-price auction both lie on the Privacy Frontier. Finally, we introduce two novel protocols that also lie on the Privacy Frontier.

Given our focus on deterministic protocols, we consider a first-price auction rule with lexicographic tie-breaking. We model type spaces as symmetric, so $\Theta_i = \Theta \subset \mathbb{R}_+$ for all $i\in N$. Let $\triangleright$ denote the lexicographic order on $\mathbb{R}_+ \times N$, with bidders ordered in decreasing priority. Let $\trimax$ denote the maximum of a set in the $\triangleright$ operator.

\begin{definition} Given strictly increasing bidding functions $b_i : \Theta_i \to \mathbb{R}_+$ for $i \in \{1, \dots, n\}$, the \emph{first-price auction choice rule} $\phi^{FPA}$ is defined by 

$$\phi^{FPA}(\bm{\theta}) = (\phi_1^{FPA},\dots,\phi_n^{FPA})(\bm{\theta}) =  ((q_1,t_1),\dots,(q_n,t_n))(\bm{\theta})$$

with

$$
(q_i,t_i)(\bm{\theta}) = \begin{cases}
    (1,b_i(\theta_i)) & \text{ if } i = \triargmax_{j\in N}\{ (b_j(\theta_j), j)\} \\
    (0, 0) & \text{ otherwise}.
\end{cases}
$$
    
\end{definition}

Defining the first-price auction allocation rule with respect to given bid functions disentangles incentive compatibility and privacy concerns. Whether $\phi^{FPA}$ is incentive-compatible depends on the choice of bid functions $\{b_i\}$, while for any set of bidding functions, the same privacy concerns persist.  We also assume that bidding functions are \textit{interleaved}: for all $i \in N$ and all $\theta_i, \theta_i' \in \Theta_i$ with $\theta_i < \theta_i'$, there exists $\hat{\bm{\theta}}_{-i} \in \bm{\Theta}_{-i}$ such that $(b_i(\theta_i), i) \triangleleft \trimax_{j\neq i} (b_j(\hat{\theta}_j), j) \triangleleft (b_i(\theta_i'), i)$. As one example, symmetric type spaces and bidding functions satisfy the interleaving assumption.\footnote{This assumption is motivated by the first-price auction: no bidder can reduce their bid without affecting the outcome at some realized type profile. This property must be assumed separately for our analysis with discrete type spaces.}

\begin{lemma}\label{lem:fpa-indistinguishable-corners-violation}
    Suppose $b_i = b$ for all $i$ and $|\Theta| \geq 4$.\footnote{The symmetry assumption ensures sufficient overlap between bidders' type spaces; weaker assumptions are also sufficient to produce violations.} Then $\phi^{FPA}$ violates the indistinguishable corners condition for every bidder.
\end{lemma}

Our analysis of the first-price auction example in \cref{fig:fpa-2v2} essentially explains why $\phi^{FPA}$ violates the indistinguishable corners condition. More generally, consider the pair of values $\theta_i,\theta_i' \in \Theta_i$ with $\theta_i < \theta_i'$. If some other bidder $j\in N$ satisfies $(b_j(\theta_j), j) \triangleright  (b_i(\theta_i'), i)$, then bidder $i$ is not pivotal to the outcome at these types: $i$ will lose the auction regardless. However, if the highest other bidder $j$ satisfies $(b_i(\theta_i), i) \triangleleft (b_j(\theta_j), j) \triangleleft (b_i(\theta_i'), i)$, then bidder $i$ is pivotal at these types: whether $i$ has value $\theta_i$ or $\theta_i'$ determines the auction's winner. Finally, if bidder $i$ loses the auction, all they observe is that they lose the auction.\footnote{This last point is why the indistinguishable corners condition is violated, despite the corners condition not being violated. The existence of a protocol that produces no contextual privacy violations (the descending auction protocol) is consistent with the absence of corners condition violations in $\phi^{FPA}$. However, because $\phi^{FPA}$ violates the indistinguishable corners condition, we may apply \cref{thm:impossibility-thm}.} Given that our impossibility theorem applies, what protocols are on the Privacy Frontier? Two natural candidates are the descending protocol and the sealed-bid protocol.

\begin{definition}[\textit{Descending auction protocol}]
   Initialize the set of possible type profiles $\prod_i \tilde{\Theta}_i$ to be $\prod_i \Theta_i$. At each stage, there exists $i \in N$ such that $(b_i(\max\tilde{\Theta}_i), i) \triangleright (b_j(\theta_j), j)$ for all $j \neq i$ and all $\theta_j \in \tilde{\Theta}_j$. Ask bidder $i$ if $\theta_i = \max \tilde{\Theta}_i$. If so, award the object to bidder $i$ at price $b_i(\max \tilde{\Theta}_i)$; otherwise, update $\tilde{\Theta}_i$ by removing $\max \tilde{\Theta}_i$ and continue.
\end{definition}

The descending auction protocol only reveals the winner's value to the auctioneer, so as \cite{brandt2008privacy} and \cite{haupt2025contextually} note, it reveals no unnecessary information to the auctioneer. However, our impossibility theorem (\cref{thm:impossibility-thm}) and \cref{lem:fpa-indistinguishable-corners-violation} then imply that the descending auction protocol reveals unnecessary information to \textit{bidders}. When a losing bidder $i$ observes that the clock has stopped at $b$, they immediately infer that the winner won at price $b$. As such, any protocol that implements $\phi^{FPA}$ reveals strictly more information to bidders than the sealed-bid first-price auction protocol, or reveals strictly more information to the auctioneer than the descending auction protocol. Despite this, the descending protocol is on the Privacy Frontier, as it reveals no unnecessary information to the auctioneer. We show the sealed-bid first-price auction is also on the Privacy Frontier, as it reveals no unnecessary information to bidders.

\begin{proposition}\label{prop:fpa-descending-and-sealed-bid-on-frontier}
    The descending auction protocol and the sealed-bid first-price auction protocol lie on the Privacy Frontier.
\end{proposition}

Although the sealed-bid first-price auction protocol reveals no unnecessary information to bidders, this does not immediately imply that it lies on the Privacy Frontier.\footnote{For example, in the second-price auction setting, the sealed-bid auction does not lie on the Privacy Frontier.} We prove \cref{prop:fpa-descending-and-sealed-bid-on-frontier} in two parts. For the descending auction, we verify both conditions of \cref{thm:suff-conditions-frontier} directly: Ex-Post Potential Pivotality follows from no unnecessary information being revealed to the auctioneer as query are potentially pivotal at the realized profile itself. No Avoidable Bidder Distinctions holds because a losing bidder's experience---the set of prices at which they declined the object---is exactly the complement of what the auctioneer learns about them. For the sealed-bid auction, we verify the conditions of \cref{prop:static-protocol-privacy-frontier-condition} (as a sealed-bid mechanism is static), which shows that revealing no unnecessary information to bidders requires the auctioneer to fully distinguish bidders' types.

\subsection{Descending Then Sealed-Bid Protocol}

We now introduce two new protocols on the Privacy Frontier that implement $\phi^{FPA}$. The \textit{descending-then-sealed-bid protocol} reveals more information to bidders when bidders have high values, but reveals more information to the auctioneer when bidders have low values. Informally, the protocol proceeds as follows:

\begin{enumerate}
    \item Run a descending auction until some predetermined threshold. 
    \item If some bidder claimed the object in stage 1, the auction has finished.
    \item Otherwise, run a sealed-bid auction.
\end{enumerate}

To illustrate, suppose there are two bidders, $A$ and $B$, each with values in $\{1, 2, 3, 4, 5\}$. The labels in \cref{fig:fpa-choice-rule-example} plot $\phi^{FPA}$, where $i, v$ denotes bidder $i$ winning at a price of $b_i(v)$. If the designer chooses a descending then sealed-bid protocol with a threshold price between $b_i(3)$ and $b_i(4)$, the partition in \cref{fig:descending-then-sealed-bid-partition} describes the information that the protocol reveals to the auctioneer, $\mathcal{P}(\sim_{(P, A)})$.\footnote{For this numerical example, to be precise, we need $\max\{b_1(3), b_2(3)\} \leq \min \{b_1(4), b_2(4)\}$.}

\begin{figure}[h]
    \centering
    \begin{subfigure}[b]{0.48\textwidth}
        \centering
        \begin{tikzpicture}[scale=1.0, thick]
        
        \useasboundingbox (0,-1) rectangle (5,5);

        \fill[red, opacity=0.5] (0,0) rectangle (1,1);   
        \fill[red, opacity=0.4] (1,0) rectangle (2,2);   
        \fill[red, opacity=0.3] (2,0) rectangle (3,3);   
        \fill[red, opacity=0.2] (3,0) rectangle (4,4);  
        \fill[red, opacity=0.1] (4,0) rectangle (5,5);   

        \fill[blue, opacity=0.4] (0,1) rectangle (1,2);   
        \fill[blue, opacity=0.3] (0,2) rectangle (2,3);   
        \fill[blue, opacity=0.2] (0,3) rectangle (3,4);   
        \fill[blue, opacity=0.1] (0,4) rectangle (4,5);  

        \draw (0,0) rectangle (5,5);

        \foreach \x in {1,...,5} {
            \node[below] at (\x-0.5,0) {\x};
        }
        \node[below=8pt] at (2.5,-0.3) {$A$};

        \foreach \y in {1,...,5} {
            \node[left] at (0,\y-0.5) {\y};
        }
        \node[left=8pt] at (-0.3,2.5) {$B$};

        \node[text=black] at (0.5,0.5) {A,1};
        \node[text=black] at (1.5,1) {A,2};
        \node[text=black] at (2.5,1.5) {A,3};
        \node[text=black] at (3.5,2) {A,4};
        \node[text=black] at (4.5,2.5) {A,5};

        \node[text=black] at (0.5,1.5) {B,2};
        \node[text=black] at (1,2.5) {B,3};
        \node[text=black] at (1.5,3.5) {B,4};
        \node[text=black] at (2,4.5) {B,5};

        \end{tikzpicture}
        \caption{Choice rule for $\phi^{FPA}$}
        \label{fig:fpa-choice-rule-example}
    \end{subfigure}
    \hfill 
    \begin{subfigure}[b]{0.48\textwidth}
        \centering
        \begin{tikzpicture}[scale=1.0, thick]

        \useasboundingbox (0,-1) rectangle (5,5);

        \fill[red, opacity=0.5] (0,0) rectangle (1,1);   
        \fill[red, opacity=0.4] (1,0) rectangle (2,2);   
        \fill[red, opacity=0.3] (2,0) rectangle (3,3);   
        \fill[red, opacity=0.2] (3,0) rectangle (4,4);  
        \fill[red, opacity=0.1] (4,0) rectangle (5,5);   

        \fill[blue, opacity=0.4] (0,1) rectangle (1,2);   
        \fill[blue, opacity=0.3] (0,2) rectangle (2,3);   
        \fill[blue, opacity=0.2] (0,3) rectangle (3,4);   
        \fill[blue, opacity=0.1] (0,4) rectangle (4,5);  

        \draw (0,0) rectangle (5,5);

        \foreach \x in {1,...,5} {
            \node[below] at (\x-0.5,0) {\x};
        }
        \node[below=8pt] at (2.5,-0.3) {$A$};

        \foreach \y in {1,...,5} {
            \node[left] at (0,\y-0.5) {\y};
        }
        \node[left=8pt] at (-0.3,2.5) {$B$};

        \node[text=black] at (3.5,2) {A,4};
        \node[text=black] at (4.5,2.5) {A,5};

        \node[text=black] at (1.5,3.5) {B,4};
        \node[text=black] at (2,4.5) {B,5};

        \node[text=black] at (0.5,0.5) {A,1};
        \node[text=black] at (1.5,1.5) {A,2};
        \node[text=black] at (1.5,0.5) {A,2};
        \node[text=black] at (0.5,1.5) {B,2};
        \node[text=black] at (2.5,1.5) {A,3};
        \node[text=black] at (2.5,0.5) {A,3};
        \node[text=black] at (2.5,2.5) {A,3};
        \node[text=black] at (1.5,2.5) {B,3};
        \node[text=black] at (0.5,2.5) {B,3};

        \draw (0,1) -- (3,1);
        \draw (0,2) -- (3,2);
        \draw (1,0) -- (1,3);
        \draw (2,0) -- (2,3);

        \draw (3,0) -- (3,4);
        \draw (4,0) -- (4,5);
        \draw (0,3) -- (3,3);
        \draw (0,4) -- (4,4);

        \end{tikzpicture}
        \caption{$\mathcal{P}(\sim_{(P,A)})$, descending then sealed-bid}
        \label{fig:descending-then-sealed-bid-partition}
    \end{subfigure}
    
    \caption{Descending then Sealed-Bid Protocol}
    \label{fig:descending-then-sealed-bid-example}
\end{figure}

Suppose the designer cares about privacy from bidders more than privacy from the auctioneer at low valuations (perhaps because post-auction competition is more intense at these valuations) but cares more about privacy from the auctioneer than privacy from bidders at high valuations (perhaps because bidders with especially high valuations in this auction are more likely to win future auctions, which the auctioneer can use to their advantage). The descending then sealed bid protocol may then be a natural choice: it reveals more information to the auctioneer about low valuations, but reveals more information to bidders in high valuations. This protocol also lies on the Privacy Frontier.

\begin{definition}
    For a threshold price $p$, let $\hat{\Theta}_j := \{\theta_j \in \Theta_j: b_j(\theta_j) \leq p\}$ and let $\hat{\bm{\Theta}} = \bigtimes_j \hat{\Theta}_j$. Define $\phi': \bm{\Theta} \to X \cup \{\hat{x}\}$ as $\phi'(\bm{\theta}) = \hat{x}$ if $\bm{\theta} \in \hat{\bm{\Theta}}$ and $\phi'(\bm{\theta}) = \phi(\bm{\theta})$ otherwise. 
    The \emph{descending then sealed-bid} protocol runs the descending protocol until the auctioneer either determines the outcome or determines $\bm{\theta} \in \hat{\bm{\Theta}}$. If $\bm{\theta} \in \hat{\bm{\Theta}}$, the auctioneer runs a sealed-bid protocol: any bidder $j$ with $|\hat{\Theta}_j| \geq 2$ reports her type, and the auctioneer chooses the outcome.
\end{definition}

\begin{proposition}\label{prop:desc-ending}
    The descending then sealed-bid protocol lies on the Privacy Frontier for $\phi^{FPA}$.
\end{proposition} 

By choosing threshold price $p$, the designer can change when the auction switches from the descending stage to the sealed-bid stage. If one descending then sealed-bid protocol descends for longer than another, then it preserves more privacy from the auctioneer at the cost of bidders learning more. At the two extremes of this tradeoff are the descending protocol and sealed-bid protocol themselves. 

\subsection{Descending-Holdout Protocol}

In some settings, it may be especially important to protect one particular agent's privacy.\footnote{For example, one bidder may have particularly sensitive data or technological information that affects their bid. As another example, one bidder may be a long-lived agent concerned about repeated interactions with third parties, while other bidders are short-lived agents bidding in only one interaction.} The \emph{descending-holdout protocol} implements $\phi^{FPA}$ with this motivation in mind by revealing less information about one chosen bidder than both the sealed-bid first-price protocol and the descending protocol. In contrast with either of these existing auction formats, the designer can now protect the chosen agent's privacy from \emph{both} the auctioneer and other bidders.

The descending-holdout protocol fixes one bidder $k$ to be held out. The auctioneer first solicits sealed bids from all bidders $j \neq k$, and selects the highest bidder as the tentative winner $i \in N$. The auctioneer then runs a descending auction for bidder $k$ with a secret reserve price of $b_i(\theta_i)$. If bidder $k$ claims the item at a price above $b_i(\theta_i)$, bidder $k$ wins the item. Otherwise, bidder $i$ wins the item.

\begin{definition}
    The \emph{descending-holdout} protocol fixes bidder $k$ as the ``holdout'' bidder. Each bidder $j \neq k$ has one information set with actions $b_j(\Theta_j)$. The tentative winner is bidder $i = \triargmax_{j \neq k} \{(b_j(\theta_j), j)\}$; let $b_i(\tilde{\theta}_i)$ denote the price they tentatively win at. 
    
    The auctioneer then sequentially asks bidder $k$ if they claim the item for a price of $b_k(\tilde{\theta}_k)$, with $\tilde{\theta}_k$ beginning at $\max \{ \Theta_k \}$ and descending in each step. The protocol ends when either $k$ claims the item at some $b_k(\tilde{\theta}_k)$ with $(b_k(\tilde{\theta}_k), k) \triangleright (b_i(\tilde{\theta}_i), i)$ or the tentative winner beats the holdout bidder in which case $(b_i(\tilde{\theta}_i), i) \triangleright (b_k(\theta_k),k)$. If $k$ claims the item, they win the item at price $b_k(\tilde{\theta}_k)$. Otherwise, bidder $i$ wins the item at price $b_i(\tilde{\theta}_i)$.
\end{definition}

\cref{fig:partition-descending-holdout} displays the partition of information the auctioneer learns, $\mathcal{P}(\sim_{P,A})$, in the descending-holdout protocol with Bidder B as the holdout bidder. We then show that the descending-holdout protocol lies on the Privacy Frontier.

\begin{figure}[h]
    \centering
    \begin{tikzpicture}[scale=1.0, thick]

    \useasboundingbox (0,-1) rectangle (5,5);

    \fill[red, opacity=0.5] (0,0) rectangle (1,1);   
    \fill[red, opacity=0.4] (1,0) rectangle (2,2);   
    \fill[red, opacity=0.3] (2,0) rectangle (3,3);   
    \fill[red, opacity=0.2] (3,0) rectangle (4,4);  
    \fill[red, opacity=0.1] (4,0) rectangle (5,5);   

    \fill[blue, opacity=0.4] (0,1) rectangle (1,2);   
    \fill[blue, opacity=0.3] (0,2) rectangle (2,3);   
    \fill[blue, opacity=0.2] (0,3) rectangle (3,4);   
    \fill[blue, opacity=0.1] (0,4) rectangle (4,5);  

    \draw (0,0) rectangle (5,5);

    \foreach \x in {1,...,5} {
        \node[below] at (\x-0.5,0) {\x};
    }
    \node[below=8pt] at (2.5,-0.3) {Bidder $A$};

    \foreach \y in {1,...,5} {
        \node[left] at (0,\y-0.5) {\y};
    }
    \node[left=8pt] at (-0.3,2.5) {Bidder $B$};

    \node[text=black] at (3.5,2) {A,4};
    \node[text=black] at (4.5,2.5) {A,5};

    \node[text=black] at (0.5,3.5) {B,4};
    \node[text=black] at (1.5,3.5) {B,4};
    \node[text=black] at (2.5,3.5) {B,4};

    \node[text=black] at (0.5,4.5) {B,5};
    \node[text=black] at (1.5,4.5) {B,5};
    \node[text=black] at (2.5,4.5) {B,5};
    \node[text=black] at (3.5,4.5) {B,5};

    \node[text=black] at (0.5,0.5) {A,1};
    \node[text=black] at (1.5,1.0) {A,2};
    \node[text=black] at (0.5,1.5) {B,2};
    \node[text=black] at (2.5,1.5) {A,3};
    \node[text=black] at (1.5,2.5) {B,3};
    \node[text=black] at (0.5,2.5) {B,3};

    \draw (1,0) -- (1,5);
    \draw (2,0) -- (2,5);
    \draw (3,0) -- (3,5);
    \draw (4,0) -- (4,5);

    \draw (0,1) -- (1,1);
    \draw (0,2) -- (2,2);
    \draw (0,3) -- (3,3);
    \draw (0,4) -- (4,4);

    \end{tikzpicture}
    \caption{Partition induced by descending-holdout protocol}
    \label{fig:partition-descending-holdout}
\end{figure}

\begin{proposition}\label{prop:descending-holdout-frontier}
    The descending-holdout protocol lies on the Privacy Frontier for $\phi^{FPA}$.
\end{proposition}

The descending-holdout protocol highlights the tradeoffs that a designer makes in choosing which bidder's information to protect. Compared to the sealed-bid protocol, the descending-holdout protocol reveals less information about the holdout bidder $k$ to the auctioneer: the auctioneer only learns bidder $k$'s exact value if $k$ wins the auction. Compared to the descending protocol, the descending-holdout protocol reveals less information about bidder $k$ to other bidders: if bidder $k$ wins the auction, other bidders still only learn that some other bidder won the auction, and in particular, they do not learn bidder $k$'s exact winning bid. In this sense, the descending-holdout protocol reveals minimal necessary information about bidder $k$ to both the auctioneer and other bidders.\footnote{In addition, the descending-holdout protocol satisfies Bayesian incentive compatibility for all bidders. For all bidders $j \neq k$, the protocol is strategically equivalent to a sealed-bid first-price auction. For bidder $k$, the protocol is strategically equivalent to a descending auction.} Our core tradeoff still holds: The descending-holdout protocol protects bidder $k$'s information from both the auctioneer and other bidders, but the auctioneer and bidder $k$ learn about \textit{other} bidders' values.

Overall, descending auctions are minimally auctioneer informative, while sealed-bid auctions are minimally bidder informative but also maximally auctioneer informative. Descending-then-sealed-bid auctions can interpolate between the two extremes based on when the sealed-bid portion of the protocol begins. Finally, descending-holdout auctions interpolate between descending and sealed-bid protocols in a different manner and cannot be ordered against descending-then-sealed-bid auctions. 

\section{Second-Price Auctions}\label{sec:second-price-auctions}

Consider an auctioneer implementing a second-price auction choice rule.\footnote{An auctioneer may wish to use a second-price choice rule to maintain ex-post incentive compatibility, for example.} In this section, we show the second-price auction choice rule violates the indistinguishable corners condition, so our impossibility theorem applies. Furthermore, our definition of the Privacy Frontier has bite: the sealed-bid second-price auction is \textit{not} on the Privacy Frontier. However, the ascending auction lies on the Privacy Frontier, and it reveals less information to bidders than the ascending-join auction of \cite{haupt2025contextually}.

\begin{definition} The \emph{second-price auction choice rule $\phi^{SPA}$} is defined by\footnote{We follow the convention of defining the winner's payment as the minimum value the winner could have reported and still won the auction. In a discrete type space, this may differ from the second-highest value.}

$$\phi^{SPA}(\bm{\theta}) = (\phi_1^{SPA},\dots,\phi_n^{SPA})(\bm{\theta}) =  ((q_1,t_1),\dots,(q_n,t_n))(\bm{\theta})$$
with
$$
q_i(\bm{\theta}) = \begin{cases}
    1 & \text{ if } i = \triargmax_{j\in N} \{ (\theta_j, j)\}, \\
    0 & \text{ otherwise}.
\end{cases}
$$
$$
t_i(\bm{\theta}) = \begin{cases}
    \min\{ \theta_i' \in \Theta_i : q_i(\theta_i', \bm{\theta}_{-i}) = 1 \} & \text{ if } q_i(\bm{\theta}) = 1, \\
    0 & \text{ otherwise}.
\end{cases}
$$
\end{definition}

We continue to assume symmetric type spaces, with $\Theta_i = \Theta$ for all $i$. Unlike the first-price auction, the second-price auction violates the standard corners condition.\footnote{Our impossibility theorem still has implications beyond \cite{haupt2025contextually}, even in settings with only violations of the standard corners condition. In such a setting, the designer can construct a protocol (in this case, the ascending-join protocol) which produces no contextual privacy violations for one protected bidder; our impossibility theorem shows that such a protocol cannot be minimally bidder-informative.} $\phi^{SPA}$ also has indistinguishable corners condition violations which are not corners condition violations. Intuitively, whenever a bidder is relevant in determining the price but not who receives the good, they become pivotal between outcomes indistinguishable to them. 

\begin{lemma}\label{lem:spa-indistinguishable-corners-violation}
    Suppose $|\Theta| \geq 5$ and $n \geq 3$. Then for each bidder, there exist type profiles that violate the indistinguishable corners condition for $\phi^{SPA}$ but do not violate the corners condition.
\end{lemma}

\cref{lem:spa-indistinguishable-corners-violation} implies that the conclusions of \cref{thm:impossibility-thm} hold, and the designer must trade off between privacy from the auctioneer and privacy from bidders. However, some natural protocols do \emph{not} lie on the Privacy Frontier: in particular, the sealed-bid second-price auction does not lie on the frontier.

One way to show this is by verifying that the conditions of \cref{prop:static-protocol-privacy-frontier-condition} do \textit{not} hold, as the characterization is sharp for static protocols. In particular, the sealed-bid second-price auction does not satisfy ex-post potential pivotality. Suppose $\Theta_i = \{1,2,3,4\}$, and bidder $i$ with value $4$ wins the auction at price $2$. As it is a sealed-bid auction, the auctioneer learns that $\theta_i = 4$; in particular, the auctioneer distinguishes type $\theta_i=4$ from type $\theta_i = 3$. But at the end of the auction, bidder $i$ knows this information is \emph{not} pivotal: if $\theta_i=3$ instead of $\theta_i=4$, then bidder $i$ would still win the object at a price of $2$.

This analysis also illustrates how a different format can improve on the privacy properties of the sealed-bid second-price auction, while not revealing more information to bidders: by revealing less information to the auctioneer about the winner's value. One protocol which privacy-dominates the sealed-bid second-price auction while implementing $\phi^{SPA}$ is the familiar \emph{ascending auction}. The ascending protocol increases the price offered to each bidder until all but one bidder drops out.

\begin{definition}[\textit{Ascending auction protocol}]
    Let $\tilde{\bm{\Theta}} = \bm{\Theta}$ be the set of possible type profiles and let $\mathcal{A}=N$ be the set of active bidders. At each stage, set $i =  \triargmin_{j\in \mathcal{A}} \{ (\min \tilde{\Theta}_j, j)\}$ and ask $i$ if $\theta_i > \min\tilde{\Theta}_i$. If so, remove $\min \tilde{\Theta}_i$ from $\tilde{\Theta}_i$; otherwise, remove $i$ from $\mathcal{A}$. If $|\mathcal{A}|=1$, let $j$ be the bidder in $\mathcal{A}$ and award the good to $j$ at price $\min \tilde{\Theta}_j$.
\end{definition}

\begin{proposition}\label{prop:spa-sealed-bid-dominated-by-ascending}
    The ascending auction privacy-dominates the sealed-bid second-price auction. Consequently, the sealed-bid second-price auction does not lie on the Privacy Frontier.
\end{proposition}

The ascending auction may initially appear to reveal more to the winning bidder than the sealed-bid auction does. However, the only additional information that the ascending auction reveals to the winner is the second-highest value. Since the winner learns this information via their payment regardless of the protocol used, the ascending auction reveals the same information to bidders as the sealed-bid auction does. 

Other protocols implement $\phi^{SPA}$ and reveal strictly less to the auctioneer than the ascending auction. One example is the ascending-join protocol of \cite{haupt2025contextually}, when $N > 2$. The ascending-join protocol begins by increasing the clocks for two bidders until one bidder drops out, then successively asks new bidders whether their value is above the current second-highest value. (We defer the formal definition of this protocol to the Appendix.) However, the ascending-join protocol reveals more information to \textit{bidders} than the ascending auction. If a bidder's first query is ``Is your value at least $p$?'' the bidder immediately learns that the second-highest value among all previous bidders is $p$. The ascending-join protocol therefore reveals unnecessary information to bidders, and does not privacy-dominate the ascending auction.\footnote{This also follows immediately from Theorem \ref{thm:impossibility-thm} by noting that there is no contextual privacy violation for the final bidder who joins in the ascending-join protocol.} Finally, we show that the ascending auction lies on the Privacy Frontier.

\begin{proposition}\label{prop:ascending-frontier}
    The ascending auction lies on the Privacy Frontier. 
\end{proposition}

The ascending auction reveals all losing bidders' values to the auctioneer, and reveals no unnecessary information about the winning bidder to the auctioneer. The key step of the proof is verifying that the ascending auction satisfies ex-post potential pivotality, which allows us to apply \cref{thm:suff-conditions-frontier}. This implies that the auctioneer must learn all losing bidders' values to avoid revealing unnecessary information to bidders. We consider a type profile $\bm{\theta}_{-i}$ and $\theta_i,\theta_i' \in \bm{\theta}_{i}$ such that bidder $i$ loses when their value is $\theta_i$ or $\theta_i'$. We construct a type profile $\bm{\theta}_{-i}'$ at which bidder $i$ is pivotal. Intuitively, when the highest other bidder has value between $\theta_i$ and $\theta_i'$, then $i$'s information is pivotal to the outcome: whether their value is $\theta_i$ or $\theta_i'$ determines the auction's winner.

Taken together, our results illustrate that a privacy-conscious designer need not prefer the ascending-join protocol: if she values privacy from bidders, the designer may instead prefer the ascending protocol. However, a privacy-conscious designer should prefer the ascending protocol to the sealed-bid second-price protocol: the sealed-bid protocol is privacy-dominated by the ascending protocol.

\section{Related Literature}\label{sec:related-literature}

Our work is related to the notions of unconditional privacy (which has the strict requirement of zero privacy violations) and contextual privacy (which only analyzes privacy from the auctioneer). \cite{brandt2008privacy} consider implementing auctions via bilateral communication among bidders without revealing any information, a condition known as \emph{unconditional privacy} (and considered earlier by \cite{chor1991boolean}). Similarly, \cite{Kushelvitz1992privacy} considers unconditional privacy when agents communicate among themselves instead of bilaterally with a mediator and additionally studies complexity of private protocols. We instead consider settings with an auctioneer and characterize a Privacy Frontier even when unconditional privacy is unattainable. \cite{milgrom2020clock} and \cite{ausubel2004auction} consider the related notion of \emph{unconditional winner privacy}, which reveals no unnecessary information about the winner's value; in contrast, we consider the privacy of all bidders. Privacy from the auctioneer is related to \textit{relative informativeness} (\cite{mackenzie2022menu}, \cite{segal2007communication}), but we further include what bidders learn from the protocol in our analysis. \cite{haupt2025contextually} characterizes protocols that implement choice rules while revealing minimal information \emph{to the auctioneer}, which they call contextual privacy. We show that contextual privacy typically conflicts with the desire to reveal minimal information \emph{to bidders}, leading to a distinct Privacy Frontier.

Other approaches to privacy include \cite{Liu2025WorstCasePrivacy}, which quantifies the worst-case privacy risk of information structures and studies optimal mechanisms subject to a ``budget'' of privacy violations. We instead study the maximally privacy-preserving implementation of a given allocation rule. \cite{mechanismdesigninformationleakage} models agents who may learn other agents' actions through channels outside of the mechanism; we study what information must be revealed by the mechanism itself.
In contrast with \cite{eilat2021bayesian}, who propose a quantitative measure of information that a mechanism reveals to the designer, we study what agents learn about each other during the mechanism. The literature on differential privacy (\cite{dwork2006differential}, \cite{pai2013privacy}) focuses on what a designer learns from a mechanism's outcome; we broaden the focus to what both the auctioneer and agents learn from playing the mechanism itself.

A broader literature considers protocols that use trusted randomization or computational hardness assumptions. See \cite{alvarez2020survey} for a survey of cryptographic protocols in auctions and see \cite{bogetoft2009sugarbeets} for applications to an agricultural auction. \cite{izmalkov2011perfect} use a trusted randomization device to implement any choice rule while fully preserving privacy. In contrast, we  provide a benchmark for what privacy protections are possible without resorting to cryptographic protocols or trusted randomization. 

Other literature discusses the importance of bidders' privacy in real-world settings, typically nesting a mechanism in a larger model.
\cite{rothkopf1990vickrey} consider bidders who are reluctant to reveal their true valuations in Vickrey auctions due to post-auction bargaining.
\cite{milgrom2020clock}, \cite{ausubel2004auction} and \cite{mcmillan1994spectrum} discuss the importance of protecting winner privacy in spectrum auctions and multi-unit auctions. \cite{dworczak2020aftermarkets} models an aftermarket which is affected by the information revealed in an auction. We are motivated by similar concerns as this literature, and we provide a framework to compare whose privacy is compromised by a given protocol, and who their information is revealed to. 

Finally, our work relates to the broader literature on extensive-form mechanism design. \cite{Segal2013Communication} utilizes extensive-form protocols to study the communication complexity of verifying that a given allocation rule is faithfully executed. 
\cite{AkbarpourLi2020CredibleAuctions}, \cite{grigoryan2023auditability}, \cite{banchio2025dynamic}, and \cite{komo2025shill} consider an auctioneer or mediator deviating from a specified mechanism. In contrast to this literature, we consider an auctioneer who faithfully executes a specified protocol. Privacy concerns persist even if the auctioneer can commit to the protocol used. Privacy may also be a component of the credibility tradeoff: Whether such deviations are profitable is informed by what the auctioneer learns, while whether such deviations are detectable is disciplined by what bidders learn. 

\section{Conclusion}\label{sec:conclusion}

We consider a designer who constructs a deterministic bilateral communication protocol among an auctioneer and multiple bidders to implement a given choice rule. Information leakages to the auctioneer or bidders may have unintended and negative consequences in a variety of economic settings. However, the designer faces an inherent tradeoff between preventing the auctioneer from unnecessarily learning bidders' private information and preventing bidders from unnecessarily learning each others' private information. Protocols can be partially ordered based on how much information they reveal to the auctioneer or to other bidders; protocols for which no alternative protocols reveal less to both the auctioneer and to other bidders lie on the \emph{Privacy Frontier}. We provide sufficient conditions for a designer to verify that a protocol lies on the Privacy Frontier.

For first-price auctions, the sealed-bid protocol (which reveals minimal information to bidders) and the descending protocol (which reveals minimal information to the auctioneer) lie on the frontier. We introduce the descending then sealed-bid protocol to trade off between these two extremes in different ranges of bidder values. We also introduce the descending-holdout protocol to protect the privacy of one selected bidder. For second-price auctions, the ascending protocol lies on the frontier. However, our privacy considerations have bite: The sealed-bid second-price auction is not on the Privacy Frontier. These tradeoffs persist beyond auction settings: our impossibility theorem holds for arbitrary choice rules. Our results highlight that the criterion of simply preserving privacy is incomplete without first answering the question of privacy \textit{from whom}. 
 
Several avenues for further research remain. Future work could utilize our sufficient conditions to establish a sharp characterization of all protocols on the frontier.
To illustrate the difficulties in characterizing the frontier, consider the effects of removing a single question from a protocol. This reduces the information revealed to the auctioneer. However, it may be infeasible if future questions condition on bidders' responses to the question. Furthermore, it may increase or decrease the information revealed to bidders, depending on whether bidders are being asked the same question for different realizations of the type profile. More generally, the space of all protocols implementing a fixed choice rule can be complex.

Future work could also connect privacy to other concerns in the extensive-form mechanism design literature. If bidders learn more along the path of play of a protocol, the protocol is less private from bidders. However, this could result in (1) bidders needing less contingent reasoning, resulting in a protocol that is strategically simpler or (2) bidders having more information to detect potential auctioneer deviations. This interplay between privacy and other economic forces is important yet underdeveloped.

\newpage

\bibliography{cites}

\appendix

\newpage

\section*{Appendix A: Pruning and Incentive Compatibility}
\makeatletter\def\@currentlabel{Appendix A}\makeatother
\label{Appendix A}

This appendix comments on the incentive properties of protocols. If the choice rule the designer seeks to implement is dominant-strategy incentive compatible, then it is without loss to restrict to protocols $P = (G,\bm{S})$ which are ex-post incentive compatible \citep{NagelSaitto2024AsIfDS}. Recall that $g(z)$ is the outcome at terminal history $z$ and $z(\bm S)$ is the terminal history reached at strategies $S$. Let $u_i: X \times \Theta_i \to \mathbf{R}$ denote bidder $i$'s utility function.

\begin{definition}[Incentive Compatibility]
     A choice rule $\phi$ is \emph{dominant-strategy incentive compatible} if it is a dominant strategy for each bidder to truthfully reveal their type: for all bidders $i$, $\theta_i, \theta_i' \in \Theta_i, \bm{\theta}_{-i} \in \bTheta_{-i}$,
    $$u_i(\phi(\theta_i, \bm{\theta}_{-i}), \theta_i) \geq u_i(\phi(\theta_i', \bm{\theta}_{-i}), \theta_i).$$
    A protocol $P = (G, \{S_i(\theta_i)\}_{i \in N})$ is \emph{ex-post incentive compatible} if, for any realization of other agents' types, each agent finds it optimal to follow the designer's strategy recommendation. That is, for all bidders $i$, $\theta_i \in \Theta_i, \bm{\theta}_{-i} \in \bm{\Theta}_{-i}, \sigma_i' \in \Sigma_i$,
    $$u_i(g(z(S_i(\theta_i), \bm S_{-i}(\bm{\theta}_{-i}))), \theta_i) \geq u_i(g(z(\sigma_i', \bm S_{-i}(\bm{\theta}_{-i}))), \theta_i).$$
\end{definition}

We will restrict attention to protocols where each history is reached at some type profile. Protocols where some histories are never reached at any type profile can be ``pruned'' to produce an alternative protocol such that the preceding condition is satisfied without changing privacy. Pruning also simplifies strategic analysis: At non-pruned protocols, deviations to histories that are never relevant may be profitable, but the designer can simply prune these branches of the game tree.

\begin{definition}[Pruning]
    Take $G = (H, i, \{\mcal I_i\}_{i \in N}, \{A_i\}_{i \in N}, A, I, g)$ and strategies $\bm S$. Then, let $\tilde{G} = (\tilde{H}, i, \{\tilde{\mcal{I}_i}\}_{i \in N}, \{\tilde{A_i}\}_{i \in N}, \tilde{A}, \tilde{I}, \tilde{g})$ be the game with:
    \begin{enumerate}
        \item $\tilde{H} = \{h \in H: \exists \bm{\theta} \in \bm{\Theta} \text{ s.t. } h \text{ is reached along the path of play of } \bm S(\bm{\theta})\}$.
        \item $\tilde{\mcal{I}}_i = \{I \cap \tilde{H}\}_{I \in \mcal I_i}$ for each $i \in N$.
        \item $(\tilde{i}, \{\tilde{A}_i\}_{i \in N}, \tilde{A}, \tilde{I})$ are $(i, \{A_i\}_{i \in N}, A, I)$ restricted to $\tilde{H}$.
    \end{enumerate}
    Given $P = (G, \bm S)$ let $\tilde{P} = (\tilde{G}, \bm S)$ denote the protocol where $G$ is pruned and $\bm S$ remains unchanged.
\end{definition}

Since $P$ implements $\phi$ on-path and $P = \tilde{P}$ for each realized path of play, $\tilde{P}$ implements $\phi$ as well. As such, it is without loss for the purpose of implementation to restrict to pruned protocols. It is also without loss for the purpose of analyzing privacy. Privacy violations are determined by on-path behavior so pruning off-path histories has no effect.

\begin{proposition}[Pruning Preserves Privacy]\label{prop:pruning-privacy}
    Let $\tilde{P} = (\tilde{G}, \bm S)$ be the protocol attained from pruning $P = (G, \bm S)$. Then:
    \begin{enumerate}
        \item Auctioneer Privacy Equivalence: $\bm{\theta} \sim_{(P, A)} \bm{\theta}'$ if and only if $\bm{\theta} \sim_{(\tilde{P}, A)} \bm{\theta}'$.
        \item Bidder Privacy Equivalence: For each bidder $i$, $\bm{\theta} \sim_{(P, i)} \bm{\theta}'$ if and only if $\bm{\theta} \sim_{(\tilde{P}, i)} \bm{\theta}'$.
    \end{enumerate}
\end{proposition}

This result is of independent interest beyond incentive compatibility. It allows the designer to simplify the search for protocols on the frontier to protocols where each history is reached at some path of play; in particular, it places an explicit bound on the maximal length of protocols which do not ask redundant questions. 

In any pruned protocol, each history is reached on-path for some type, so deviations in the game $G$ are essentially deviations where a bidder pretends to be a different type. As such, incentive compatibility of the choice rule is able to rule out all possible deviations. 

\begin{proposition}[Incentive Compatibility of Pruned Protocols]\label{prop:pruning-IC}
    Suppose $P = (G, \bm S)$ is pruned and implements $\phi$. If $\phi$ is dominant-strategy incentive compatible, then $P$ is ex-post incentive compatible.
\end{proposition}

Protocols do not necessarily preserve Bayesian incentive compatibility, however. A protocol could provide certain bidders with pertinent information \textit{before} they make a choice. As such, interim beliefs may be different under the protocol implementing the mechanism compared to its normal form game, so Bayesian incentive compatibility does not translate. This may provide another novel justification for dominant strategy incentive compatibility: Dominant strategy incentive compatibility is robust to the details of how a mechanism is actually run whereas Bayesian incentive compatibility is not. Furthermore, our impossibility theorem only gets stronger if Bayesian incentive compatibility is required as it restricts the set of potential protocols the designer may use. In our application to first-price auctions, for which the choice rule $\phi^{FPA}$ is not dominant-strategy incentive compatible, we show that certain auctions are not privacy-dominated by any protocol implementing $\phi^{FPA}$. This immediately implies that these auctions are not privacy-dominated by any incentive-compatible protocol implementing $\phi^{FPA}$. For additional discussion, see \cite{FadelSegal2009CommunicationCost} which investigates the communication costs of requiring incentive compatibility in extensive-form protocols. 

\newpage

\section*{Appendix B: Non-Necessity of Theorem \ref{thm:suff-conditions-frontier}}
\makeatletter\def\@currentlabel{Appendix B}\makeatother
\label{Appendix B}

This Appendix provides two examples. Each example consists of a choice function and a protocol on the Privacy Frontier for that choice function. In the first example, the protocol does not satisfy Ex-Post Potential Pivotality. In the second example, the protocol does not satisfy No Avoidable Bidder Distinctions. Furthermore, these two examples can be joined together (by taking the product of the type spaces, outcomes, choice rules, and protocols) to produce a choice function and a protocol on the Privacy Frontier for that choice function that neither satisfies Ex-Post Potential Pivotality nor No Avoidable Bidder Distinctions.

\subsection*{Non-Necessity of Ex-Post Potential Pivotality}

Let $\Theta_1 = \{a, b\}, \Theta_2 = \{x, y, z\}$, and $X = \{1, 2, 3\}$ with no outcomes distinguishable from one another. Let $\phi$ be as in the following figure: 

\begin{figure}[h!]
    \centering
    \begin{tikzpicture}[scale=1]
        
        \draw[thick] (0,0) rectangle (6,4);
        
        \draw[thin] (2,0) -- (2,4);
        \draw[thin] (4,0) -- (4,4);
        
        \draw[thin] (0,2) -- (6,2);
        
        \node at (1,4.6) {$x$};
        \node at (3,4.6) {$y$};
        \node at (5,4.6) {$z$};
        
        \node[left] at (-0.4,3) {$a$};
        \node[left] at (-0.4,1) {$b$};
        
        \node at (1,3) {1};
        \node at (3,3) {3};
        \node at (5,3) {3};
        
        \node at (1,1) {2};
        \node at (3,1) {3};
        \node at (5,1) {3};
        
    \end{tikzpicture}
    \caption{Allocation Rule $\phi$}
    \label{ACROP_not_nec_allocation}
\end{figure}

Consider the following protocol: 
\begin{enumerate}
    \item Ask Bidder two to fully reveal their type. 
    \item If Bidder two has type $x$ or $y$, ask bidder one to fully reveal their type. 
\end{enumerate}
This induces the following partition: 
\begin{figure}[h!]
    \centering
    \begin{tikzpicture}[scale=1]
        
        \draw[thick] (0,0) rectangle (6,4);
        
        \draw[line width=1.6pt] (0, 2) -- (4, 2);
        \draw[line width=1.6pt] (2,0) -- (2,4);
        \draw[line width=1.6pt] (4,0) -- (4,4);

        \node at (1,4.6) {$x$};
        \node at (3,4.6) {$y$};
        \node at (5,4.6) {$z$};
        
        \node[left] at (-0.4,3) {$a$};
        \node[left] at (-0.4,1) {$b$};
        
        \node at (1,3) {1};
        \node at (3,3) {3};
        \node at (5,3) {3};
        
        \node at (1,1) {2};
        \node at (3,1) {3};
        \node at (5,1) {3};
        
    \end{tikzpicture}
    \caption{Partition Induced by Protocol On Frontier Violating Ex-Post Potential Pivotality}
    \label{ACROP_not_nec_protocol}
\end{figure}

\newpage

This protocol is on the frontier: No alternative protocol revealing strictly less to the auctioneer asks bidder one if their type is $a$ or $b$ when bidder two's type is $y$ (since no longer asking bidder one this question is the only way to reveal less information to the auctioneer while still implementing $\phi$). As such, any protocol that reveals strictly less to the auctioneer must reveal bidder two's type to bidder one whenever bidder two's type is $x$. Thus, no protocol that reveals strictly less to the auctioneer also reveals weakly less to the bidders. Any protocol which reveals weakly but not strictly less to the auctioneer induces the same partition, so no protocol can privacy-dominate the given protocol. 

Next, this protocol violates Ex-Post Potential Pivotality. To show this, we need to find $i \in N$,  $\theta_i, \theta_i' \in \Theta_i$ and  $\bm{\theta}_{-i} \in \bm{\Theta}_{-i}$ with $(\theta_i, \bm{\theta}_{-i}) \nsim_{(P, A)} (\theta_i', \bm{\theta}_{-i})$ such that for all $\bm{\theta}_{-i}' \in \bm{\Theta}_{-i}$ at least one of the following does not hold:
\begin{enumerate}[(a)]
    \item $\phi(\theta_i, \bm{\theta}_{-i}') \neq \phi(\theta_i', \bm{\theta}_{-i}')$,
    \item $(\theta_i, \bm{\theta}_{-i}) \sim_{(P, i)} (\theta_i, \bm{\theta}_{-i}')$ or $(\theta_i', \bm{\theta}_{-i}) \sim_{(P, i)} (\theta_i', \bm{\theta}_{-i}')$. 
\end{enumerate}
Take $i = 2, \theta_i = y, \theta_i' = z, \bm{\theta}_{-i} = b$. Then $(y, b) \not \sim_{(P, A)} (z, b)$ as desired. However, whether agent $i$ has type $y$ or $z$ is never pivotal. For all $\bm{\theta}_{-i}' \in \bTheta_{-i} = \{a,b\}$ we have $\phi((y, \bm{\theta}_{-i}')) = 3 = \phi((z, \bm{\theta}_{-i}'))$, so condition (a) never holds.

\subsection*{Non-Necessity of No Avoidable Bidder Distinctions}

Let $\Theta_1 = \{a, b, c\}, \Theta_2 = \{x, y, z\}$, and $X = \{1, 2, 3, 4, 5, 6, 7\}$ with no outcomes distinguishable from one another. Let $\phi$ be as in the following figure: 

\begin{figure}[h!]
    \centering
    \begin{tikzpicture}[scale=1]
        
        \draw[thick] (0,0) rectangle (6,6);
        
        \draw[thick] (2,0) -- (2,6);
        \draw[thick] (4,0) -- (4,6);
        
        \draw[thick] (0,2) -- (6,2);
        \draw[thick] (0,4) -- (6,4);
        
        \node at (1,6.6) {$x$};
        \node at (3,6.6) {$y$};
        \node at (5,6.6) {$z$};
        
        \node[left] at (-0.4,5) {$a$};
        \node[left] at (-0.4,3) {$b$};
        \node[left] at (-0.4,1) {$c$};
        
        \node at (1,5) {1};
        \node at (3,5) {4};
        \node at (5,5) {6};
        
        \node at (1,3) {2};
        \node at (3,3) {5};
        \node at (5,3) {6};
        
        \node at (1,1) {3};
        \node at (3,1) {5};
        \node at (5,1) {7};
        
    \end{tikzpicture}
    \caption{Allocation Rule $\phi$}
    \label{NABDnot_nec_allocation}
\end{figure}

Consider the following class of protocols: 
\begin{enumerate}
    \item Ask bidder two to fully reveal their type.
    \item If bidder two's type is $x$, ask bidder one to fully reveal their type by either:
    \begin{itemize}
        \item First ask if their type is $a$; if not then ask if their type is $b$. 
        \item First ask if their type is $c$; if not then ask if their type is $b$. 
    \end{itemize}
    \item If bidder two's type is $y$, ask bidder one if their type is $a$ or not. 
    \item If bidder two's type is $z$, ask bidder one if their type is $c$ or not.
\end{enumerate}

All protocols of the above form induce the following partition: 

\begin{figure}[h!]
    \centering
    \begin{tikzpicture}[scale=1]
        
        \draw[thick] (0,0) rectangle (6,6);

        \draw[line width=1.6pt] (0,4) -- (4,4); 
        \draw[line width=1.6pt] (4,0) -- (4,6); 
        \draw[line width=1.6pt] (4,2) -- (6,2); 
        \draw[line width=1.6pt] (2,0) -- (2,6);
        \draw[line width=1.6pt] (0,2) -- (2,2);

        \node at (1,6.6) {$x$};
        \node at (3,6.6) {$y$};
        \node at (5,6.6) {$z$};
        
        \node[left] at (-0.4,5) {$a$};
        \node[left] at (-0.4,3) {$b$};
        \node[left] at (-0.4,1) {$c$};
        
        \node at (1,5) {1};
        \node at (3,5) {4};
        \node at (5,5) {6};
        
        \node at (1,3) {2};
        \node at (3,3) {5};
        \node at (5,3) {6};
        
        \node at (1,1) {3};
        \node at (3,1) {5};
        \node at (5,1) {7};
        
    \end{tikzpicture}
    \caption{Partition Induced by Frontier Protocol Violating No Avoidable Bidder Distinctions}
    \label{NABDnot_nec_protocol}
\end{figure}

No protocol can reveal strictly less to the auctioneer than a protocol in the above class as otherwise the auctioneer could not implement $\phi$. Similarly, every protocol that reveals weakly less to the auctioneer is essentially in that class of protocols already: Bidder two always reveals their type and the questions bidder one is asked are forced given that. Thus, every protocol in the above class is on the frontier. 

Then, no protocol in this class satisfies No Avoidable Bidder Distinctions. Towards a contradiction, suppose some protocol in this class did satisfy No Avoidable Bidder Distinctions for bidder one. That is, taking $i = 1$, for all $\theta_i \in \Theta_i$ and all $\bm{\theta}_{-i}, \bm{\theta}_{-i}' \in \bm{\Theta}_{-i}$ with $(\theta_i, \bm{\theta}_{-i}) \nsim_{(P, i)} (\theta_i, \bm{\theta}_{-i}')$, at least one of the following holds:
\begin{enumerate}[(a)]
    \item $\phi(\theta_i, \bm{\theta}_{-i}) \nsim_{\Omega_i} \phi(\theta_i, \bm{\theta}_{-i}')$
    \item there exists $\theta_i' \in \Theta_i$ such that $(\theta_i, \bm{\theta}_{-i}) \sim_{(P,A)} (\theta_i', \bm{\theta}_{-i})$ and $(\theta_i, \bm{\theta}_{-i}') \nsim_{(P,A)} (\theta_i', \bm{\theta}_{-i}')$
    \item there exists $\theta_i' \in \Theta_i$ such that $(\theta_i, \bm{\theta}_{-i}') \sim_{(P,A)} (\theta_i', \bm{\theta}_{-i}')$ and $(\theta_i, \bm{\theta}_{-i}) \nsim_{(P,A)} (\theta_i', \bm{\theta}_{-i})$.
\end{enumerate} 
Since no outcomes are distinguishable, it must be that either (b) or (c) holds.

First, set $\theta_i = a, \bm{\theta}_{-i} = x, \bm{\theta}_{-i}' = y$. Since $a$ is distinguished from both $b$ and $c$ at both $x$ and $y$, neither point (b) nor (c) can hold, which means that for a potential protocol to satisfy No Avoidable Bidder Distinctions, it must be that $(a, x) \sim_{(P, i)} (a, y)$. For this to be the case, the first question bidder one is asked when bidder two's type is $x$ must be the same as the first question bidder one is asked when bidder two's type is $y$, namely whether or not bidder one's type is $a$.

However, next set  $\theta_i = c, \bm{\theta}_{-i} = x, \bm{\theta}_{-i}' = z$. Since $c$ is distinguished from both $a$ and $b$ at both $x$ and $z$, neither point (b) nor (c) can hold, which means that for a potential protocol to satisfy No Avoidable Bidder Distinctions, it must be that $(c, x) \sim_{(P, i)} (c, z)$. For this to be the case, the first question bidder one is asked when bidder two's type is $x$ must be the same as the first question bidder one is asked when bidder two's type is $z$, namely whether or not bidder one's type is $c$.

This gives the desired contradiction: The first question bidder one is asked cannot simultaneously be whether or not their type is $a$ and whether or not their type is $c$. 

\section*{Appendix C: Omitted Proofs}
\makeatletter\def\@currentlabel{Appendix C}\makeatother
\label{Appendix C}

\subsection*{Privacy}

\subsubsection*{PROOF OF THEOREM \ref{thm:impossibility-thm}}

\begin{proof}
    Let $P$ be a protocol implementing choice rule $\phi$. Consider any types $\theta_i, \theta_i' \in \Theta_i$ and $\bm{\theta}_{-i}, \bm{\theta}_{-i}' \in \bm{\Theta}_{-i}$ with $\phi(\theta_i, \bm{\theta}_{-i}') \neq \phi(\theta_i', \bm{\theta}_{-i}')$. Suppose $(\theta_i, \bm{\theta}_{-i}) \sim_{(P, A)} (\theta_i', \bm{\theta}_{-i})$. Then, $\phi(\theta_i, \bm{\theta}_{-i}') \neq \phi(\theta_i', \bm{\theta}_{-i}')$ implies that for $P$ to implement $\phi$, it must be that $(\theta_i, \bm{\theta}_{-i}') \not\sim_{(P, A)} (\theta_i', \bm{\theta}_{-i}')$. By Lemma \ref{lem:auctioneer-distinguishes-both-rows}, it must be that $(\theta_i,\bm{\theta}_{-i}) \nsim_{(P, i)} (\theta_i,\bm{\theta}_{-i}')$ and $(\theta_i',\bm{\theta}_{-i}) \nsim_{(P, i)} (\theta_i',\bm{\theta}_{-i}')$ as desired. The second statement of \cref{thm:impossibility-thm} follows directly from the definitions of revealing unnecessary information. 
\end{proof}

\subsubsection*{PROOF OF LEMMA \ref{lem:auctioneer-distinguishes-both-rows}}

\begin{proof}
    Because $(\theta_i,\bm{\theta}_{-i}') \nsim_{(P, A)} (\theta_i',\bm{\theta}_{-i}')$, there exists an information set $I_i$ on the path of play for $(\theta_i,\bm{\theta}_{-i}')$ at which $S_i(\theta_i)(I_i) \neq S_i(\theta_i')(I_i)$. Because $(\theta_i,\bm{\theta}_{-i}) \sim_{(P, A)} (\theta_i',\bm{\theta}_{-i})$, at every information set $I_i$ on the path of play for $(\theta_i,\bm{\theta}_{-i})$, we have $S_i(\theta_i)(I_i) = S_i(\theta_i')(I_i)$. Hence there exists some information set $I_i$ on the path of play for $(\theta_i,\bm{\theta}_{-i}')$ which is not on the path of play for $(\theta_i,\bm{\theta}_{-i})$, and therefore $o_i(\theta_i,\bm{\theta}_{-i}) \neq o_i(\theta_i,\bm{\theta}_{-i}')$. By definition, $(\theta_i,\bm{\theta}_{-i}) \nsim_{(P, i)} (\theta_i,\bm{\theta}_{-i}')$. By an identical argument, there exists some information set on the path of play for $(\theta_i',\bm{\theta}_{-i}')$ which is not on the path of play for $(\theta_i',\bm{\theta}_{-i})$, and therefore $(\theta_i',\bm{\theta}_{-i}) \nsim_{(P, i)} (\theta_i',\bm{\theta}_{-i}')$. The second sentence in the lemma follows from the contrapositive.
\end{proof}

\subsubsection*{PROOF OF THEOREM \ref{thm:suff-conditions-frontier}}

\begin{lemma}\label{lem:rectangle-property} (Rectangle property.)
    For any protocol $P$, each cell of the partition $\mathcal{P}(\sim_{(P,A)})$  is a product set.
\end{lemma}

\begin{proof}
A profile reaches terminal history $z$ if and only if, at every information set on the path to $z$, the player called there takes the action on the path. This is a condition on each $\theta_i$ separately, which yields the product form. The auctioneer's observation determines $z$, so cells in $\mathcal{P}(\sim_{(P,A)})$ are exactly these product sets.
\end{proof}

\begin{lemma}\label{lem:auctioneer-distinguishes-one-agent}
    Let $P$ and $\hat{P}$ be two protocols. Suppose that for all $i \in N$, all $\theta_i, \theta_i' \in \Theta_i$ and all $\bm{\theta}_{-i} \in \bm{\Theta}_{-i}$, $(\theta_i, \bm{\theta}_{-i}) \sim_{(P, A)} (\theta_i', \bm{\theta}_{-i})$ if and only if $(\theta_i, \bm{\theta}_{-i}) \sim_{(\hat{P}, A)} (\theta_i', \bm{\theta}_{-i})$. Then $\sim_{(P, A)}$ and $\sim_{(\hat{P}, A)}$ are identical.
\end{lemma}

\begin{proof}
    Consider any $\bm{\theta}, \btheta' \in \bTheta$.
    There exists a sequence of type profiles \\ $\{ \bm{\theta}^{(1)}, \bm{\theta}^{(2)}, \dots, \bm{\theta}^{(K)} \}$ which transforms $\bm{\theta}$ into $\bm{\theta}'$ by altering one agent's type in each step. That is, $\bm{\theta}^{(1)} = \bm{\theta}$ and $\bm{\theta}^{(K)} = \bm{\theta}'$, where for all $k \in \{1,\dots, K-1\}$ there exists agent $i(k) \in N$ and $\theta_{i(k)}, \theta_{i(k)}' \in \Theta_{i(k)}$ such that $\bm{\theta}^{(k)} = (\theta_{i(k)}, \bm{\theta}^{(k)}_{-i(k)})$ and $\bm{\theta}^{(k+1)} = (\theta_{i(k)}', \bm{\theta}^{(k)}_{-i(k)})$. In particular, we may take a sequence for which each $i(k)$ is distinct: the sequence alters each agent's type at most once.

    If $\bm{\theta}^{(k)} \sim_{(P,A)} \bm{\theta}^{(k+1)}$ for all $k \in \{1,\dots, K-1\}$ then by transitivity, $\bm{\theta} \sim_{(P,A)} \bm{\theta}'$.

    If $\bm{\theta} \sim_{(P,A)} \bm{\theta}'$, then by the rectangle property (\cref{lem:rectangle-property}), $\bm{\theta}^{(k)} \sim_{(P,A)} \bm{\theta}^{(k+1)}$ for all $k$, because all intermediate type profiles lie in the same cell. By contrapositive, if $\bm{\theta}^{(k)} \nsim_{(P,A)} \bm{\theta}^{(k+1)}$ for some $k \in \{1,\dots, K-1\}$, then $\bm{\theta} \nsim_{(P,A)} \bm{\theta}'$.

    The same logic applies to $\hat{P}$. Because $\sim_{(P, A)}$ and $\sim_{(\hat{P}, A)}$ coincide on the comparisons of type profiles $\bm{\theta}^{(k)},\bm{\theta}^{(k+1)}$ for all $k$, $\sim_{(P, A)}$ and $\sim_{(\hat{P}, A)}$ are therefore identical.
\end{proof}

\begin{proof}
    The proof proceeds as follows. Consider any protocol $P'$ that (weakly) privacy-dominates $P$ and implements $\phi$. First, we show that if $P$ satisfies Ex-Post Potential Pivotality, then $P'$ must reveal the same information to the auctioneer as $P$ (Lemma \ref{lem:Ex-Post Potential Pivotality}). Second, if $P$ satisfies No Avoidable Bidder Distinctions, then if $P'$ reveals the same information to the auctioneer as $P$, it must also reveal the same information to bidders as $P$ (Lemma \ref{lem:No Avoidable Bidder Distinctions}). 

    Putting the two lemmas together, suppose $P$ satisfies Ex-Post Potential Pivotality and No Avoidable Bidder Distinctions. Consider any protocol $\hat{P}$ (weakly) privacy-dominating $P$. By Ex-Post Potential Pivotality and Lemma \ref{lem:Ex-Post Potential Pivotality}, $\sim_{(\hat{P}, A)}$ is identical to $\sim_{(P, A)}$. Then, by No Avoidable Bidder Distinctions and Lemma \ref{lem:No Avoidable Bidder Distinctions}, $\sim_{(\hat{P}, i)}$ is identical to $\sim_{(P, i)}$ for all $i \in N$. Hence no protocol $\hat{P}$ can strictly privacy-dominate $P$, and $P$ lies on the Privacy Frontier for $\phi$.

    We now state and prove the two lemmas. 

    \begin{lemma}\label{lem:Ex-Post Potential Pivotality}
        Suppose $P$ satisfies Ex-Post Potential Pivotality. If protocol $\hat{P}$ privacy-dominates $P$, then $\sim_{(\hat{P}, A)}$ is identical to $\sim_{(P, A)}$.
    \end{lemma}

    \begin{proof}
        The proof proceeds in two steps. First, we will show that if $\hat{P}$ privacy-dominates $P$, then $\sim_{(P, A)}$ and $\sim_{(\hat{P}, A)}$ are identical for any pair of type profiles that differ in only one agent's type. Second, we can extend this pairwise result to a global equivalence result. 

        Suppose $P$ satisfies Ex-Post Potential Pivotality and $\hat{P}$ privacy-dominates $P$. As $P$ satisfies Ex-Post Potential Pivotality, for all $i \in N$, all $\theta_i, \theta_i' \in \Theta_i$ and all $\bm{\theta}_{-i} \in \bm{\Theta}_{-i}$, if $(\theta_i, \bm{\theta}_{-i}) \nsim_{(P, A)} (\theta_i', \bm{\theta}_{-i})$, at least one of the following holds:
        \begin{enumerate}[(a)]
            \item $\phi(\theta_i, \bm{\theta}_{-i}) \neq \phi(\theta_i', \bm{\theta}_{-i})$
            \item there exists $\bm{\theta}_{-i}' \in \bm{\Theta}_{-i}$ such that $\phi(\theta_i, \bm{\theta}_{-i}') \neq \phi(\theta_i', \bm{\theta}_{-i}')$ and $(\theta_i, \bm{\theta}_{-i}) \sim_{(P, i)} (\theta_i, \bm{\theta}_{-i}')$ 
            \item there exists $\bm{\theta}_{-i}' \in \bm{\Theta}_{-i}$ such that $\phi(\theta_i, \bm{\theta}_{-i}') \neq \phi(\theta_i', \bm{\theta}_{-i}')$ and $(\theta_i', \bm{\theta}_{-i}) \sim_{(P, i)} (\theta_i', \bm{\theta}_{-i}')$.
        \end{enumerate}
        Then, $\hat{P}$ privacy-dominates $P$, so $(\theta_i, \bm{\theta}_{-i}) \sim_{(P, A)} (\theta_i', \bm{\theta}_{-i})$ implies $(\theta_i, \bm{\theta}_{-i}) \sim_{(\hat{P}, A)} (\theta_i', \bm{\theta}_{-i})$. Hence it suffices to show that $(\theta_i, \bm{\theta}_{-i}) \nsim_{(P, A)} (\theta_i', \bm{\theta}_{-i})$ implies $(\theta_i, \bm{\theta}_{-i}) \nsim_{(\hat{P}, A)} (\theta_i', \bm{\theta}_{-i})$. There are three cases corresponding to which of the conditions in the Lemma statement hold:
        \begin{itemize}
            \item In case (a), $(\theta_i, \bm{\theta}_{-i}) \nsim_{(\hat{P}, A)} (\theta_i', \bm{\theta}_{-i})$ because $\hat{P}$ implements $\phi$.
            \item In case (b), $(\theta_i, \bm{\theta}_{-i}') \nsim_{(\hat{P}, A)} (\theta_i', \bm{\theta}_{-i}')$ because $\hat{P}$ implements $\phi$ and $\phi(\theta_i, \bm{\theta}_{-i}') \neq \phi(\theta_i', \bm{\theta}_{-i}')$. Furthermore, $(\theta_i, \bm{\theta}_{-i}) \sim_{(\hat{P}, i)} (\theta_i, \bm{\theta}_{-i}')$ because $\hat{P}$ privacy-dominates $P$. By \cref{lem:auctioneer-distinguishes-both-rows}, $(\theta_i, \bm{\theta}_{-i}) \nsim_{(\hat{P}, A)} (\theta_i', \bm{\theta}_{-i})$. 
            \item Case (c) is identical to case (b) with the roles of $\theta_i$ and $\theta_i'$ reversed.
        \end{itemize}
        Thus, $(\theta_i, \bm{\theta}_{-i}) \sim_{(P, A)} (\theta_i', \bm{\theta}_{-i})$ if and only if $(\theta_i, \bm{\theta}_{-i}) \sim_{(\hat{P}, A)} (\theta_i', \bm{\theta}_{-i})$. By \cref{lem:auctioneer-distinguishes-one-agent}, $\sim_{(P, A)}$ is identical to $\sim_{(\hat{P}, A)}$.
    \end{proof}

    \begin{lemma}\label{lem:No Avoidable Bidder Distinctions}
        Suppose $P$ satisfies No Avoidable Bidder Distinctions. If protocol $\hat{P}$ privacy-dominates $P$ and $\sim_{(\hat{P}, A)}$ is identical to $\sim_{(P, A)}$, then $\sim_{(\hat{P}, i)}$ is identical to $\sim_{(P, i)}$.
    \end{lemma}

    \begin{proof}
        Suppose $P$ satisfies No Avoidable Bidder Distinctions. Then, taking the contrapositive of the conditions given in the definition, for all $\theta_i \in \Theta_i$ and all $\bm{\theta}_{-i}, \bm{\theta}_{-i}' \in \bm{\Theta}_{-i}$, if $(\theta_i, \bm{\theta}_{-i}) \nsim_{(P, i)} (\theta_i, \bm{\theta}_{-i}')$, at least one of the following holds:

        \begin{enumerate}[(a)]
            \item $\phi(\theta_i, \bm{\theta}_{-i}) \nsim_{\Omega_i} \phi(\theta_i, \bm{\theta}_{-i}')$
            \item there exists $\theta_i' \in \Theta_i$ such that $(\theta_i, \bm{\theta}_{-i}) \sim_{(P,A)} (\theta_i', \bm{\theta}_{-i})$ and $(\theta_i, \bm{\theta}_{-i}') \nsim_{(P,A)} (\theta_i', \bm{\theta}_{-i}')$
            \item there exists $\theta_i' \in \Theta_i$ such that $(\theta_i, \bm{\theta}_{-i}') \sim_{(P,A)} (\theta_i', \bm{\theta}_{-i}')$ and $(\theta_i, \bm{\theta}_{-i}) \nsim_{(P,A)} (\theta_i', \bm{\theta}_{-i})$.
        \end{enumerate}

        Under this re-writing of the definition of No Avoidable Bidder Distinctions, the proof of \cref{lem:No Avoidable Bidder Distinctions} follows similarly to the proof of Lemma \ref{lem:Ex-Post Potential Pivotality}: A fixed set of auctioneer violations pins down a particular set of bidder violations.

        Suppose $\hat{P}$ privacy-dominates $P$, so $(\theta_i, \bm{\theta}_{-i}) \sim_{(P, i)} (\theta_i, \bm{\theta}_{-i}')$ implies $(\theta_i, \bm{\theta}_{-i}) \sim_{(\hat{P}, i)} (\theta_i, \bm{\theta}_{-i}')$. Hence it suffices to show that $(\theta_i, \bm{\theta}_{-i}) \nsim_{(P, i)} (\theta_i, \bm{\theta}_{-i}')$ implies $(\theta_i, \bm{\theta}_{-i}) \nsim_{(\hat{P}, i)} (\theta_i, \bm{\theta}_{-i}')$.\footnote{In the other case, if any two type profiles $\bm{\theta}$ and $\bm{\theta}'$ have $\theta_i \neq \theta_i'$, then $\bm{\theta} \nsim_{(P,i)} \bm{\theta}'$ and $\bm{\theta} \nsim_{(\hat{P},i)} \bm{\theta}'$, because $i$ has different types in the two profiles.}
        
        In case (a), $(\theta_i, \bm{\theta}_{-i}) \nsim_{(\hat{P}, i)} (\theta_i, \bm{\theta}_{-i}')$ because bidder $i$ can distinguish the two distinct outcomes. In case (b), \cref{lem:auctioneer-distinguishes-both-rows} implies $(\theta_i, \bm{\theta}_{-i}) \nsim_{(\hat{P}, i)} (\theta_i, \bm{\theta}_{-i}')$. Case (c) is identical to case (b), with $\bm{\theta}_{-i}$ and $\bm{\theta}_{-i}'$ swapped.
    \end{proof}

    Lemmas \ref{lem:Ex-Post Potential Pivotality} and \ref{lem:No Avoidable Bidder Distinctions} along with the preceding discussion imply Theorem \ref{thm:suff-conditions-frontier}.
\end{proof}

We also show that revealing no unnecessary information to the auctioneer implies ex-post potential pivotality. This is convenient to later apply to the descending auction and the ascending auction protocol of \cref{sec:second-price-auctions}.

\begin{lemma}\label{lem:no-unnecessary-implies-Ex-Post Potential Pivotality}
    Let $P$ be any protocol implementing a choice rule $\phi$. If $P$ reveals no unnecessary information to the auctioneer, then $P$ satisfies Ex-Post Potential Pivotality.
\end{lemma}

\begin{proof}
    Fix $i \in N$, $\theta_i, \theta_i' \in \Theta_i$, and $\bm{\theta}_{-i} \in \bm{\Theta}_{-i}$ with $(\theta_i, \bm{\theta}_{-i}) \nsim_{(P, A)} (\theta_i', \bm{\theta}_{-i})$. Because $P$ reveals no unnecessary information to the auctioneer, the contrapositive of that definition gives $\phi(\theta_i, \bm{\theta}_{-i}) \neq \phi(\theta_i', \bm{\theta}_{-i})$. Take $\bm{\theta}_{-i}' := \bm{\theta}_{-i}$: condition (b) of Ex-Post Potential Pivotality is the preceding display, and condition (a) holds because $\sim_{(P, i)}$ is reflexive.
\end{proof}

In words, when the auctioneer learns nothing beyond the outcome, every query is pivotal at the realized profile itself, so no counterfactual profile is needed. Combining the descending protocol producing no auctioneer privacy violations with \cref{lem:no-unnecessary-implies-Ex-Post Potential Pivotality} gives that $P^{\mathrm{D}}$ satisfies Ex-Post Potential Pivotality.

\begin{lemma}\label{lem:sealed-bid-implies-no-avoidable-bidder-distinctions}
    Let $P$ be a static protocol. Then $P$ satisfies no avoidable bidder distinctions.
\end{lemma}

\begin{proof}
    In a static protocol $P$, each bidder has exactly one information set in their experience. Hence $(\theta_i, \bm{\theta}_{-i}) \sim_{(P, i)} (\theta_i, \bm{\theta}_{-i}')$ if and only if $\phi(\theta_i, \bm{\theta}_{-i}) \sim_{\Omega_i} \phi(\theta_i, \bm{\theta}_{-i}')$, which implies that no avoidable bidder distinctions is satisfied.
\end{proof}

\subsubsection*{PROOF OF PROPOSITION \ref{prop:static-protocol-privacy-frontier-condition}}

\begin{proof}
  (\textbf{$\Leftarrow$}) By \cref{lem:sealed-bid-implies-no-avoidable-bidder-distinctions} and \cref{thm:suff-conditions-frontier}, $P$ lies on the Privacy Frontier.

  (\textbf{$\Rightarrow$}) The overall idea is that if $P$ does not satisfy Ex-Post Potential Pivotality, we may feasibly remove questions from $P$ while maintaining a protocol that reveals no unnecessary information to bidders.

  The baseline information revealed by $P$ is as follows.
  Let $\mathcal{M}_i$ be the partition of $\Theta_i$ into classes of types that send the same message, and let $[\theta_i]$ be the respective class of $\theta_i$. The information revealed to the auctioneer is the product partition $\Pi_i \mathcal{M}_i$. By implementation, $\phi$ is constant on each element of this partition and $\phi([\bm{\theta}])$ denotes the outcome on element $[\btheta] \in \Pi_i \mathcal{M}_i$. $P$ reveals no unnecessary information to bidders: $\bm{\theta} \sim_{(P, i)} \bm{\theta}'$ if and only if $\phi(\bm{\theta}) \sim_{\Omega_i} \phi(\bm{\theta'})$ and $\theta_i = \theta_i'$, as the protocol is static. 

  Suppose $P$ does not satisfy Ex-Post Potential Pivotality. Then there exists agent $i \in N$, types $\theta_i, \theta_i' \in \Theta_i$ with $[\theta_i] \neq [\theta_i']$, and $\btheta_{-i} \in \bm{\Theta}_{-i}$ such that $\phi(\theta_i, \btheta_{-i}) = \phi(\theta_i', \btheta_{-i})$ and for all $\btheta_{-i}' \in \bm{\Theta}_{-i}$, at least one of the following holds:

  \begin{enumerate}
    \item[(a)] $\phi(\theta_i, \bm{\theta}_{-i}') = \phi(\theta_i', \bm{\theta}_{-i}')$,
    \item[(b)] $(\theta_i, \bm{\theta}_{-i}) \nsim_{(P, i)} (\theta_i, \bm{\theta}_{-i}')$ and $(\theta_i', \bm{\theta}_{-i}) \nsim_{(P, i)} (\theta_i', \bm{\theta}_{-i}')$.
  \end{enumerate}

  Fix such a $\theta_i, \theta_i' \in \Theta_i$ and $\btheta_{-i} \in \bm{\Theta}_{-i}$. Let $\tilde{\bm{\Theta}}_{-i} := \{\bm{\theta}_{-i}' \in \bm{\Theta}_{-i}: (\theta_i, \bm{\theta}_{-i}) \sim_{(P, i)} (\theta_i, \bm{\theta}_{-i}') \text{ or } (\theta_i', \bm{\theta}_{-i}) \sim_{(P, i)} (\theta_i', \bm{\theta}_{-i}')\}$ be the set of $\bm{\theta}_{-i}'$ that do not satisfy condition (b) above. In particular, for all $\btheta_{-i}' \in \tilde{\bm{\Theta}}_{-i}$, we must have $\phi(\theta_i, \bm{\theta}_{-i}') = \phi(\theta_i', \bm{\theta}_{-i}')$. Further, $\tilde{\bm{\Theta}}_{-i}$ is nonempty because $\bm{\theta}_{-i} \in \tilde{\bm{\Theta}}_{-i}$.\footnote{This also implies that for any $\btheta_{-i}' \in \tilde{\bm{\Theta}}_{-i}$, both $(\theta_i, \bm{\theta}_{-i}) \sim_{(P, i)} (\theta_i, \bm{\theta}_{-i}') \textit{ and } (\theta_i', \bm{\theta}_{-i}) \sim_{(P, i)} (\theta_i', \bm{\theta}_{-i}')$. This is because by (a) we have $\phi(\theta_i, \btheta_{-i}) = \phi(\theta_i', \btheta_{-i})$ and $\phi(\theta_i, \btheta_{-i}') = \phi(\theta_i',\btheta_{-i}')$, hence $(\theta_i, \bm{\theta}_{-i}) \sim_{(P, i)} (\theta_i, \bm{\theta}_{-i}')$ if and only if $(\theta_i', \bm{\theta}_{-i}) \sim_{(P, i)} (\theta_i', \bm{\theta}_{-i}')$.} Note that $\tilde{\bTheta}_{-i}$ is a union of elements of $\Pi_{j\neq i} \mathcal{M}_j$, as $i$ cannot distinguish two type profiles that are in the same element of $\Pi_{j\neq i} \mathcal{M}_j$.

  We construct a new protocol $\hat{P}$ as follows:

  \begin{enumerate}
    \item Each agent $j \neq i$ has one information set and sends message $[\theta_j]$. This yields an element $[\btheta_{-i}]'$ of $\Pi_{j\neq i} \mathcal{M}_j$.
    \item Ask agent $i$ if her type is in $[\theta_i] \cup [\theta_i']$.
    \begin{enumerate}
      \item If $i$'s type is not in $[\theta_i] \cup [\theta_i']$, ask $i$ to report her type class from $\mathcal{M}_i$.
      \item If $i$'s type is in $[\theta_i] \cup [\theta_i']$ and $[\btheta_{-i}]' \nsubseteq \tilde{\bTheta}_{-i}$, ask agent $i$ to report whether her type is in $[\theta_i]$ or $[\theta_i']$.
    \end{enumerate}
    \item If agent $i$ reported her exact type class, the outcome is $\phi([\bm{\theta}])$ from agents' reports. Otherwise, agent $i$'s reported type is in $[\theta_i] \cup [\theta_i']$ and $[\btheta_{-i}]' \subseteq \tilde{\bTheta}_{-i}$, and the outcome is $\phi([\theta_i], [\btheta_{-i}]')$.
  \end{enumerate}

  Protocol $\hat{P}$ implements the choice rule because $\phi(\theta_i, \bm{\theta}_{-i}') = \phi(\theta_i', \bm{\theta}_{-i}')$ for all $\bm{\theta}_{-i}' \in \tilde{\bm{\Theta}}_{-i}$. $\hat{P}$ reveals strictly less information to the auctioneer than $P$, because the auctioneer sometimes does not learn if agent $i$ has type in $[\theta_i]$ or $[\theta_i']$. 
  
  Finally, $\hat{P}$ still reveals no unnecessary information to bidders. An agent $j \neq i$ still only learns what she observes of the outcome, because $j$ has only one information set. For agent $i$, if her type is not in $[\theta_i] \cup [\theta_i']$, she observes the same sequence of play regardless of other agents' types $\bm{\theta}_{-i}$, hence learns nothing besides the outcome. If agent $i$'s type is in $[\theta_i] \cup [\theta_i']$, the only additional types she could possibly distinguish are distinguishing some $\btheta_{-i}' \in \tilde{\bTheta}_{-i}$ from some $\btheta_{-i}'' \notin \tilde{\bTheta}_{-i}$. But for all such type profiles:

  $$
  \phi([\theta_i], [\btheta_{-i}]') \sim_{\Omega_i} \phi([\theta_i], [\btheta_{-i}]) \nsim_{\Omega_i} \phi([\theta_i], [\btheta_{-i}]'')
  $$

  and hence agent $i$ already distinguished these pairs in $P$. In the same way, $\phi([\theta_i'], [\btheta_{-i}]') \sim_{\Omega_i} \phi([\theta_i'], [\btheta_{-i}]) \nsim_{\Omega_i} \phi([\theta_i'], [\btheta_{-i}]'')$. Hence $\hat{P}$ reveals the same information to bidders as $P$ and reveals strictly less information to the auctioneer, so $P$ is privacy-dominated.

\end{proof}

\subsection*{First-Price Auctions}

\subsubsection*{PROOF OF LEMMA \ref{lem:fpa-indistinguishable-corners-violation}}

\begin{proof}
    Consider any two bidders $i,j$. Choose $\theta_i<\theta_j'<\theta_i'<\theta_j$ and some $\bm{\theta}_{-ij}$ with $\theta_k \leq \theta_i$ for all $k\neq i,j$. Let $\bm{\theta}_{-i} = (\theta_j,\bm{\theta}_{-ij})$ and $\bm{\theta}_{-i}' = (\theta_j',\bm{\theta}_{-ij})$.
    
    Then $\phi(\theta_i,\bm{\theta}_{-i}) = \phi(\theta_i',\bm{\theta}_{-i})$ because both cases lead to bidder $j$ winning at their bid $b_j(\theta_j)$. We also have $\phi(\theta_i,\bm{\theta}_{-i}) \sim_{\Omega_i} \phi(\theta_i,\bm{\theta}_{-i}')$ because bidder $i$ loses the auction in both cases. Finally, $\phi(\theta_i,\bm{\theta}_{-i}') \neq \phi(\theta_i',\bm{\theta}_{-i}')$, because bidder $i$ loses the auction in the first case and wins the auction in the second case. 
\end{proof}

\subsubsection*{PROOF OF PROPOSITION \ref{prop:fpa-descending-and-sealed-bid-on-frontier} (DESCENDING AUCTION)}

\begin{proof}
    
We prove that the descending auction protocol, denoted $P^{\mathrm{D}}$, lies on the Privacy Frontier by verifying the two conditions of \cref{thm:suff-conditions-frontier}. Ex-Post Potential Pivotality follows from \cref{lem:no-unnecessary-implies-Ex-Post Potential Pivotality}: any protocol revealing no unnecessary information to the auctioneer satisfies Ex-Post Potential Pivotality. No Avoidable Bidder Distinctions uses the structure of the descending auction and is established in \cref{lem:descending-No Avoidable Bidder Distinctions}. 

For each $\bm{\theta} \in \bm{\Theta}$, let $\beta^*(\bm{\theta}) := \trimax_{j \in N}\{(b_j(\theta_j), j)\}$ denote the \emph{winning pair} and let $Q_i(\bm{\theta}) := \{v \in \Theta_i : (b_i(v), i) \triangleright \beta^*(\bm{\theta})\}$ denote the set of values about which bidder $i$ is queried before the auction ends. 

The descending protocol reveals no unnecessary information to the auctioneer: $\btheta \sim_{(P^D, A)} \btheta'$ if and only if $\phi^{FPA}(\btheta) = \phi^{FPA}(\btheta')$. This is because the protocol asks about the pairs $(b_i(v), i)$ in decreasing $\triangleright$ order and stops at the first yes, which occurs at the winning pair $\beta^*(\btheta)$. The auctioneer's history is therefore a function of $\beta^*(\btheta)$ alone; that is, it is a function of $\phi^{FPA}(\btheta)$ alone. Hence if $i$ wins the auction then $\Pi_i^{P^{\mathrm{D}}}(\theta_i, \bm{\theta}_{-i}) = \{\theta_i\}$ (the auctioneer learns $i$'s value); if $i$ loses the auction then $\Pi_i^{P^{\mathrm{D}}}(\theta_i, \bm{\theta}_{-i}) = \Theta_i \setminus Q_i(\theta_i, \bm{\theta}_{-i})$ (the auctioneer only learns $i$'s value is too low to win).

\begin{lemma}\label{lem:descending-No Avoidable Bidder Distinctions}
    The descending auction protocol satisfies No Avoidable Bidder Distinctions.
\end{lemma}

\begin{proof}
    Fix $i \in N$, $\theta_i \in \Theta_i$, and $\bm{\theta}_{-i}, \bm{\theta}_{-i}' \in \bm{\Theta}_{-i}$ satisfying the hypotheses of \cref{def:no-avoidable-bidder-distinctions}, namely
    \begin{enumerate}[(H1)]
        \item $\phi^{FPA}(\theta_i, \bm{\theta}_{-i}) \sim_{\Omega_i} \phi^{FPA}(\theta_i, \bm{\theta}_{-i}')$;
        \item $\Pi_i^{P^{\mathrm{D}}}(\theta_i, \bm{\theta}_{-i}) = \Pi_i^{P^{\mathrm{D}}}(\theta_i, \bm{\theta}_{-i}')$.
    \end{enumerate}
    We must show $(\theta_i, \bm{\theta}_{-i}) \sim_{(P^{\mathrm{D}}, i)} (\theta_i, \bm{\theta}_{-i}')$. Bidder $i$ wins at $(\theta_i, \bm{\theta}_{-i})$ if and only if they win at $(\theta_i, \bm{\theta}_{-i}')$: were bidder $i$ to win at one and lose at the other, their allocation--transfer pair would be $(1, b_i(\theta_i))$ at the former and $(0,0)$ at the latter, contradicting (H1). Two cases remain.
    \begin{enumerate}
        \item \emph{Bidder $i$ wins at both profiles.} Their experience consists of their type $\theta_i$, the fixed nodes querying their values $v > \theta_i$ each answered ``no'', the fixed node querying $\theta_i$ answered ``yes'', and the observation $\omega_i \ni (1, b_i(\theta_i))$. Every component depends only on $\theta_i$ and on fixed nodes of the tree, so $(\theta_i, \bm{\theta}_{-i}) \sim_{(P^{\mathrm{D}}, i)} (\theta_i, \bm{\theta}_{-i}')$.
        \item \emph{Bidder $i$ loses at both profiles.} By the argument above, $Q_i(\theta_i, \bm{\theta}_{-i}) = \Theta_i \setminus \Pi_i^{P^{\mathrm{D}}}(\theta_i, \bm{\theta}_{-i})$ and likewise at $\bm{\theta}_{-i}'$, so (H2) yields $Q_i(\theta_i, \bm{\theta}_{-i}) = Q_i(\theta_i, \bm{\theta}_{-i}')$. In both type profiles, bidder $i$ is queried about the type profiles in $Q_i$ in descending order, answering ``no'' to each query, and finally observes $\omega_i \ni (0, 0)$; these experiences are identical. Hence $(\theta_i, \bm{\theta}_{-i}) \sim_{(P^{\mathrm{D}}, i)} (\theta_i, \bm{\theta}_{-i}')$.
    \end{enumerate}
\end{proof}

Thus, $P^{\mathrm{D}}$ satisfies both Ex-Post Potential Pivotality and No Avoidable Bidder Distinctions. \cref{thm:suff-conditions-frontier} implies that $P^{\mathrm{D}}$ lies on the Privacy Frontier.
\end{proof}

\subsubsection*{PROOF OF PROPOSITION \ref{prop:fpa-descending-and-sealed-bid-on-frontier} (SEALED-BID AUCTION)}
\begin{proof}

    In the sealed-bid first-price auction, $\btheta \sim_{(P^{SB},A)} \btheta'$ if and only if $\btheta = \btheta'$, as it is a sealed-bid protocol. Furthermore, as it is a sealed-bid protocol, $(\theta_i,\btheta_{-i}) \sim_{(P^{SB},i)} (\theta_i,\btheta_{-i}')$ if and only if $\phi(\theta_i,\btheta_{-i}) \sim_{\Omega_i} \phi(\theta_i,\btheta_{-i}')$.

    By \cref{thm:suff-conditions-frontier} it suffices to verify Ex-Post Potential Pivotality and No Avoidable Bidder Distinctions. By \cref{lem:sealed-bid-implies-no-avoidable-bidder-distinctions}, the protocol satisfies No Avoidable Bidder Distinctions. So it suffices to verify Ex-Post Potential Pivotality.

    The conditions of Ex-Post Potential Pivotality must be verified for every $i \in N$, every pair $\theta_i \neq \theta_i'$ in $\Theta_i$, and every $\btheta_{-i} \in \bTheta_{-i}$.\footnote{If $|\Theta_i| = 1$ there is nothing to verify for bidder $i$.} If $\phi(\theta_i,\bm{\theta}_{-i}) \neq \phi(\theta_i',\bm{\theta}_{-i})$, then any protocol implementing $\phi$ satisfies this condition. Hence it suffices to show this condition holds when $\phi(\theta_i,\bm{\theta}_{-i}) = \phi(\theta_i',\bm{\theta}_{-i})$. To do so, we will construct a violation of the indistinguishable corners condition using this type profile.

    Without loss of generality, suppose $\theta_i < \theta_i'$. For any type profiles with $\phi^{FPA}(\theta_i,\bm{\theta}_{-i}) = \phi^{FPA}(\theta_i',\bm{\theta}_{-i})$, we must have that bidder $i$ loses the auction in both cases (otherwise, their payment would differ in the two cases). By our interleaving assumption, we may construct type profile $\bm{\theta}_{-i}' \in \bm{\Theta}_{-i}$ such that 
    $$
    (b_i(\theta_i), i)  \triangleleft \trimax_{j \neq i} \{(b_j(\theta_j'), j)\} \triangleleft (b_i(\theta_i'), i).
    $$

    Then $\phi(\theta_i,\bm{\theta}_{-i}')$ also results in $i$ losing the auction, and hence $\phi(\theta_i,\bm{\theta}_{-i}) \sim_{\Omega_i}\phi(\theta_i,\bm{\theta}_{-i}')$. However, $\phi(\theta_i',\bm{\theta}_{-i}')$ results in $i$ winning the auction, and hence $\phi(\theta_i,\bm{\theta}_{-i}') \neq \phi(\theta_i',\bm{\theta}_{-i}')$. We conclude that the sealed-bid first-price auction satisfies Ex-Post Potential Pivotality. 
\end{proof}

\subsubsection*{PROOF OF PROPOSITION \ref{prop:desc-ending}}
\begin{proof}
    As described in the text, the descending-then-sealed-bid protocol first runs the descending protocol on choice problem $\phi'$; if this protocol outputs $\hat{x}$, it then runs the sealed-bid protocol on $\hat{\bm{\Theta}}$. Let $P$ denote this protocol. By \cref{thm:suff-conditions-frontier}, it suffices to verify Ex-Post Potential Pivotality and No Avoidable Bidder Distinctions.

    \begin{lemma}\label{lem:desc-ending1}
        $P$ satisfies Ex-Post Potential Pivotality.
    \end{lemma}

    \begin{proof}
        Suppose $(\theta_i, \bm{\theta}_{-i}) \nsim_{(P, A)}(\theta_i', \bm{\theta}_{-i})$. If $(\theta_i, \bm{\theta}_{-i}) \notin \hat{\bm{\Theta}}$ or $(\theta_i', \bm{\theta}_{-i}) \notin \hat{\bm{\Theta}}$, then $\phi^{FPA}(\theta_i, \bm{\theta}_{-i}) \neq \phi^{FPA}(\theta_i', \bm{\theta}_{-i})$: in the descending auction, the auctioneer learns only what they need to know, so Ex-Post Potential Pivotality is satisfied in this case.
    
        Otherwise, $(\theta_i, \bm{\theta}_{-i}), (\theta_i', \bm{\theta}_{-i}) \in \hat{\bm{\Theta}}$. We apply the same techniques from the proof of \cref{prop:fpa-descending-and-sealed-bid-on-frontier}. Without loss of generality, suppose $\theta_i < \theta_i'$. If $\phi^{FPA}((\theta_i, \bm{\theta}_{-i})) \neq \phi^{FPA}((\theta_i', \bm{\theta}_{-i}))$, we are done. Otherwise, bidder $i$ loses the auction in both cases.\footnote{If bidder $i$ wins the auction in both cases, their payment differs because $b_i$ is strictly increasing, and hence the outcome differs.} We may then construct a type profile $\bm{\theta}_{-i}' \in \hat{\bm{\Theta}}_{-i}$ such that 
        $$
        (b_i(\theta_i), i) \triangleleft \trimax_{j \neq i} \{(b_j(\theta_j'), j)\} \triangleleft (b_i(\theta_i'), i).
        $$
    
        Then $\phi^{FPA}(\theta_i,\bm{\theta}_{-i}')$ also results in $i$ losing the auction with the same experience ($i$ rejects all descending offers then submits a sealed bid), so $\phi^{FPA}(\theta_i,\bm{\theta}_{-i}) \sim_{\Omega_i} \phi^{FPA}(\theta_i,\bm{\theta}_{-i}')$. However, $\phi^{FPA}(\theta_i',\bm{\theta}_{-i}')$ results in $i$ winning the auction, so $\phi^{FPA}(\theta_i,\bm{\theta}_{-i}') \neq \phi^{FPA}(\theta_i',\bm{\theta}_{-i}')$. We conclude that $P$ satisfies Ex-Post Potential Pivotality.
    \end{proof}

    \begin{lemma}\label{lem:desc-ending2}
        $P$ satisfies No Avoidable Bidder Distinctions.
    \end{lemma}

    \begin{proof}
        We show the contrapositive. Suppose $(\theta_i, \bm{\theta}_{-i}) \nsim_{(P, i)} (\theta_i, \bm{\theta}_{-i}')$. There are three cases to consider.
    
        \begin{enumerate}
            \item \textit{Case 1}: $(\theta_i, \bm{\theta}_{-i}), (\theta_i, \bm{\theta}_{-i}') \in \hat{\bm{\Theta}}$. Because $P$ is minimally bidder informative in $\hat{\bm{\Theta}}$, $\phi(\theta_i, \bm{\theta}_{-i}) \nsim_{\Omega_i} \phi(\theta_i, \bm{\theta}_{-i}')$. Hence $(\theta_i, \bm{\theta}_{-i}) \nsim_{(P, i)} (\theta_i, \bm{\theta}_{-i}')$.
            
            \item \textit{Case 2}: $(\theta_i, \bm{\theta}_{-i}) \in \hat{\bm{\Theta}}$ and $(\theta_i, \bm{\theta}_{-i}') \notin \hat{\bm{\Theta}}$. Consider any other $\theta_i' \in \hat{\Theta}_i$ with $\theta_i \neq \theta_i'$.\footnote{If $|\hat{\Theta}_i| = 1$, the conditions of No Avoidable Bidder Distinctions are only met when $i$ loses the auction in both $(\theta_i, \bm{\theta}_{-i})$ and $(\theta_i, \bm{\theta}_{-i}')$ and the auctioneer fully distinguishes agent $i$'s type under both profiles. Then $(\theta_i, \bm{\theta}_{-i}) \sim_{(P, i)} (\theta_i, \bm{\theta}_{-i}')$: the clock simply descends to the end for $i$ in both cases.} Then $(\theta_i',\bm{\theta}_{-i}) \in \hat{\bm{\Theta}}$ so $(\theta_i,\bm{\theta}_{-i}) \nsim_{(P, A)}(\theta_i',\bm{\theta}_{-i})$. However, $(\theta_i,\bm{\theta}_{-i}') \sim_{(P, A)}(\theta_i',\bm{\theta}_{-i}')$, because an agent other than $i$ won the object in the descending stage. Hence $\Pi_i^P(\theta_i, \bm{\theta}_{-i}) \neq \Pi_i^P(\theta_i, \bm{\theta}_{-i}')$.
                        
            \item \textit{Case 3}: $(\theta_i, \bm{\theta}_{-i}), (\theta_i, \bm{\theta}_{-i}') \notin \hat{\bm{\Theta}}$. If $\phi(\theta_i, \bm{\theta}_{-i}) \nsim_{\Omega_i} \phi(\theta_i, \bm{\theta}_{-i}')$, we are done. Otherwise, agent $i$ loses the auction in both cases. Furthermore, there exists $\theta_i'$ such that (without loss of generality in re-labeling $\bm{\theta}_{-i}$, $\bm{\theta}_{-i}')$, bidder $i$ would have won if their type was $\theta_i'$:\footnote{Since $(\theta_i, \bm{\theta}_{-i}) \nsim_{(P, i)} (\theta_i, \bm{\theta}_{-i}')$, bidder $i$ is offered the object at some price at least once in at least one of the type profiles.}
            $$
            \trimax_{j\neq i} \{ (b_j(\theta_j),j)\} \triangleleft (b_i(\theta_i'),i)\triangleleft \trimax_{j\neq i} \{ (b_j(\theta_j'),j)\}.
            $$
            Then $(\theta_i, \bm{\theta}_{-i}') \sim_{(P, A)} (\theta_i', \bm{\theta}_{-i}')$, as agent $i$ loses the auction in both cases, but $(\theta_i, \bm{\theta}_{-i}) \nsim_{(P, A)} (\theta_i', \bm{\theta}_{-i})$, as agent $i$ wins in one case and loses in the other case. Hence $\Pi_i^P(\theta_i, \bm{\theta}_{-i}) \neq \Pi_i^P(\theta_i, \bm{\theta}_{-i}')$.
            \end{enumerate}
    \end{proof}

    By \cref{thm:suff-conditions-frontier}, $P$ lies on the Privacy Frontier.
\end{proof}

\subsubsection*{PROOF OF PROPOSITION \ref{prop:descending-holdout-frontier}}

\begin{proof}
    Let $P$ denote the descending-holdout protocol, with $k$ denoting the holdout bidder. We use $\phi$ to refer to $\phi^{FPA}$. By \cref{thm:suff-conditions-frontier} it suffices to check that $P$ satisfies Ex-Post Potential Pivotality and No Avoidable Bidder Distinctions.

    \textbf{Claim 1:} $P$ satisfies Ex-Post Potential Pivotality.
    
    For bidder $k$, if $(\theta_k, \bm{\theta}_{-k}) \nsim_{(P, A)} (\theta_k', \bm{\theta}_{-k})$ then $\phi(\theta_k, \bm{\theta}_{-k}) \neq \phi(\theta_k', \bm{\theta}_{-k})$. This is because if $k$ wins in both cases, the price they pay is different. If $k$ loses the auction in both cases, the auctioneer only learns that $k$ loses the auction.

    For bidder $i \neq k$, the auctioneer perfectly learns $\theta_i$ in protocol $P$. The logic is then similar to that of the sealed-bid first-price auction. Without loss of generality, suppose $\theta_i < \theta_i'$. If $\phi(\theta_i, \bm{\theta}_{-i}) \neq \phi(\theta_i', \bm{\theta}_{-i})$, we are done. Otherwise, bidder $i$ loses the auction in both cases.\footnote{If bidder $i$ wins the auction in both cases, their payment differs because $b_i$ is strictly increasing, and hence the outcome differs.} We may then construct a type profile $\bm{\theta}_{-i}' \in \bm{\Theta}_{-i}$ such that 

    $$
    (b_i(\theta_i), i) \triangleleft \trimax_{j \neq i} \{(b_j(\theta_j'), j)\} \triangleleft (b_i(\theta_i'), i).
    $$

    Then $\phi(\theta_i,\bm{\theta}_{-i}')$ also results in $i$ losing the auction, so $\phi(\theta_i,\bm{\theta}_{-i}) \sim_{\Omega_i} \phi(\theta_i,\bm{\theta}_{-i}')$. However, $\phi^{FPA}(\theta_i',\bm{\theta}_{-i}')$ results in $i$ winning the auction, so $\phi(\theta_i,\bm{\theta}_{-i}') \neq \phi(\theta_i',\bm{\theta}_{-i}')$.

    \textbf{Claim 2:} $P$ satisfies No Avoidable Bidder Distinctions.

    We show the contrapositive.

    For bidder $i \neq k$, bidder $i$ learns nothing from the protocol except the outcome (he has only one information set), so if $(\theta_i, \bm{\theta}_{-i}) \nsim_{(P, i)} (\theta_i, \bm{\theta}_{-i}')$ we have $\phi(\theta_i, \bm{\theta}_{-i}) \nsim_{\Omega_i} \phi(\theta_i, \bm{\theta}_{-i}')$.

    For the holdout bidder $i = k$, suppose $(\theta_i, \bm{\theta}_{-i}) \nsim_{(P, i)} (\theta_i,\bm{\theta}_{-i}')$. Let $j := \argmax_{l\neq i} \{(b_l(\theta_l), l)\}$ and $j' := \argmax_{l\neq i} \{(b_l(\theta_l'), l)\}$ denote the highest other bidders in $\bm{\theta}_{-i}$ and $\bm{\theta}_{-i}'$, respectively. Then $(b_j(\theta_j), j) \neq (b_{j'}(\theta_{j'}'), j')$, and without loss of generality we may consider $ (b_j(\theta_j), j) \triangleleft (b_{j'}(\theta_{j'}'), j')$. There are three cases.
    \begin{enumerate}
        \item If $(b_{j'}(\theta_{j'}'),j') \triangleleft (b_i(\theta_i), i)$ then $(\theta_i, \bm{\theta}_{-i}) \sim_{(P, i)} (\theta_i,\bm{\theta}_{-i}')$, because agent $i$ wins the item in both cases before reaching the reserve price. 
        \item  If $ (b_j(\theta_j), j) \triangleleft (b_i(\theta_i), i) \triangleleft (b_{j'}(\theta_{j'}'), j') $ then $\phi(\theta_i, \bm{\theta}_{-i}) \nsim_{\Omega_i} \phi(\theta_i, \bm{\theta}_{-i}')$, because agent $i$ wins in $(\theta_i, \bm{\theta}_{-i})$ and loses in $(\theta_i, \bm{\theta}_{-i}')$. 
        \item The last case is if $(b_i(\theta_i), i) \triangleleft (b_j(\theta_j), j)$. Because  $(\theta_i, \bm{\theta}_{-i}) \nsim_{(P, i)} (\theta_i,\bm{\theta}_{-i}')$, bidder $i$ receives a different set of questions in $(\theta_i, \bm{\theta}_{-i})$ and $(\theta_i,\bm{\theta}_{-i}')$. That is, the auction ends later in $(\theta_i, \bm{\theta}_{-i})$ than in $(\theta_i,\bm{\theta}_{-i}')$: $i$ received more queries in $(\theta_i, \bm{\theta}_{-i})$ than in $(\theta_i,\bm{\theta}_{-i}')$. Hence there exists $\theta_i' \in \Theta_i$ such that
        $$(b_j(\theta_j), j)\triangleleft (b_i(\theta_i'), i) \triangleleft (b_{j'}(\theta_{j'}'), j').$$
        We have $(\theta_i, \bm{\theta}_{-i}') \sim_{(P, A)} (\theta_i', \bm{\theta}_{-i}')$, because bidder $i$ loses in both cases, so the auction ends before the auctioneer distinguishes $\theta_i$ from $\theta_i'$. However, $(\theta_i, \bm{\theta}_{-i}) \nsim_{(P, A)} (\theta_i', \bm{\theta}_{-i})$, because bidder $i$ loses the auction with value $\theta_i$ and wins the auction with value $\theta_i'$. Hence $\Pi_i^P(\theta_i, \bm{\theta}_{-i}) \neq \Pi_i^P(\theta_i, \bm{\theta}_{-i}')$.
    \end{enumerate}

    By \cref{thm:suff-conditions-frontier}, $P$ lies on the Privacy Frontier.
\end{proof}

\subsection*{Second-Price Auctions}

\subsubsection*{PROOF OF LEMMA \ref{lem:spa-indistinguishable-corners-violation}}

\begin{proof}
    We will construct an indistinguishable corners condition violation for bidder $i$. Consider bidder $k := \min\{N \setminus \{i\}\}$ and bidder $j := \min\{N \setminus \{i, k\}\}$. Let $\theta_i < \theta_j' < \theta_i' < \theta_j < \theta_k$. Choose $\bm{\theta}_{-ijk}$ with $\theta_\ell \leq \theta_i$ for all $\ell \neq i, j, k$. Thus, bidder $k$ always wins the auction but their payment depends on bidders $i$ and $j$. Take $\bm{\theta}_{-i} = (\theta_j, \theta_k, \bm{\theta}_{-ijk})$ and $\bm{\theta}_{-i}' = (\theta_j', \theta_k, \bm{\theta}_{-ijk})$.

    Then, $\phi(\theta_i, \bm{\theta}_{-i}) = \phi(\theta_i', \bm{\theta}_{-i})$ as both output bidder $k$ winning at a price of $\theta_j$. Next, $\phi(\theta_i, \bm{\theta}_{-i}) \sim_{\Omega_i} \phi(\theta_i, \bm{\theta}_{-i}')$ while $\phi(\theta_i, \bm{\theta}_{-i}) \neq \phi(\theta_i, \bm{\theta}_{-i}')$ as bidder $i$ loses in both cases but the final price is different ($\theta_j$ versus $\theta_j'$). Finally, $\phi(\theta_i, \bm{\theta}_{-i}') \neq \phi(\theta_i', \bm{\theta}_{-i}')$ as the final price is different ($\theta_j'$ versus at least $\theta_i'$). As such, $\phi^{SPA}$ has an indistinguishable corners condition violation for bidder $i$ at $\{\theta_i, \theta_i'\} \times \{\bm{\theta}_{-i}, \bm{\theta}_{-i}'\}$ but does not have a corners condition violation for bidder $i$ at those type profiles.
\end{proof}

\subsubsection*{PROOF OF PROPOSITION \ref{prop:spa-sealed-bid-dominated-by-ascending}}

\begin{proof}
    In the ascending auction, the auctioneer learns all losers' values but does not learn the winner's value. Hence the ascending auction reveals strictly less to the auctioneer than the sealed-bid second-price auction. In both formats, a losing bidder learns only that they lost the auction. In both protocols, the winning bidder learns the second-highest bidder's value based on their payment. Hence both protocols reveal the same information to bidders (that is, they are both minimally bidder-informative), and the ascending auction privacy-dominates the sealed-bid second-price auction.  
\end{proof}

\subsubsection*{PROOF OF PROPOSITION \ref{prop:ascending-frontier}}

\begin{proof}
    
    For brevity, in this proof we use $\phi$ to refer to $\phi^{SPA}$ and use $P$ to refer to the ascending auction protocol.

    \textbf{Claim 1}: $P$ satisfies Ex-Post Potential Pivotality.
    
    Consider any distinct $\theta_i,\theta_i'\in \Theta_i$ and any $\bm{\theta}_{-i} \in \bm{\Theta}_{-i}$. Without loss of generality, consider $\theta_i <\theta_i'$. If $(\theta_i, i) \triangleright \trimax_{j\neq i} \{ (\theta_j, j)\}$ then $(\theta_i,\bm{\theta}_{-i}) \sim_{(P, A)} (\theta_i',\bm{\theta}_{-i})$. If only $(\theta_i', i) \triangleright  \trimax_{j\neq i} \{ (\theta_j, j)\}$, then $\phi(\theta_i, \bm{\theta}_{-i}) \neq \phi(\theta_i', \bm{\theta}_{-i})$. 
    
    Otherwise,  $(\theta_i', i) \triangleleft  \trimax_{j\neq i} \{ (\theta_j, j)\}$. Consider any $\bm{\theta}_{-i}'$ satisfying $(\theta_i, i) \triangleleft \trimax_{j\neq i} (j, \theta_j') \triangleleft (\theta_i', i)$. Then bidder $i$ loses the auction at type profiles $(\theta_i,\bm{\theta}_{-i})$ and $(\theta_i,\bm{\theta}_{-i}')$ and  $(\theta_i,\bm{\theta}_{-i}) \sim_{(P, i)}(\theta_i,\bm{\theta}_{-i}')$, as they observe the same experience in both cases. However, bidder $i$ wins the auction at type profile $(\theta_i',\bm{\theta}_{-i}')$, hence $\phi(\theta_i,\bm{\theta}_{-i}') \neq \phi(\theta_i',\bm{\theta}_{-i}')$.    
    
    \textbf{Claim 2}: $P$ satisfies No Avoidable Bidder Distinctions.

    A losing bidder observes an identical experience at any type profile in which they lose: They observe the sequence of questions up to the price at which they drop out. Hence if $\phi(\theta_i, \bm{\theta}_{-i}) \sim_{\Omega_i} \phi(\theta_i, \bm{\theta}_{-i}')$, a losing bidder has the same experiences and so $(\theta_i, \bm{\theta}_{-i}) \sim_{(P, i)} (\theta_i, \bm{\theta}_{-i}')$.

    For a winning bidder $i$, if $\phi(\theta_i, \bm{\theta}_{-i}) \sim_{\Omega_i} \phi(\theta_i, \bm{\theta}_{-i}')$ then $i$ wins at the same price $p$ in both cases. Again, $i$ has the same experience in both profiles, as they are queried up to the same price before the auction ends, so $(\theta_i, \bm{\theta}_{-i}) \sim_{(P, i)} (\theta_i, \bm{\theta}_{-i}')$.

    By \cref{thm:suff-conditions-frontier}, we conclude that $P$ lies on the Privacy Frontier.
\end{proof}

\subsubsection*{Definition of Ascending-Join Protocol (\cite{haupt2025contextually})}

Let $\tilde{\bm{\Theta}} \subset \bm{\Theta}$ denote the set of potential types remaining. Order $X$ lexicographically by price then winner identity and let $x = \min \{\phi^{SPA}(\bm{\theta}):\bm{\theta} \in \tilde{\bm{\Theta}} \}$ denote the \textit{tentative outcome}. Given $\tilde{\bm{\Theta}}$, query agents in order $i = 1, 2,..., n$ by asking: ``Can you rule out outcome $x$ with your type information?''. Formally, the query separates $\hat{\Theta}_i = \{ \tilde{\theta}_i \in \tilde{\Theta}_i : \forall \; \tilde{\bm{\theta}}_{-i} \in \tilde{\bm{\Theta}}_{-i}, \; \phi(\theta_i,\tilde{\bm{\theta}}_{-i}) \neq x\}$ from $\tilde{{\Theta}}_i \setminus \hat{\Theta}_i$. If agent $i$ reports $\theta_i \in \hat{\Theta}_i$, then $\tilde{\bm{\Theta}}$ is updated and the procedure begins again from bidder 1. If no bidder can rule out outcome $x$, then the protocol terminates and chooses outcome $x$.

\subsection*{Pruning and Incentive-Compatibility}

\subsubsection*{PROOF OF PROPOSITION \ref{prop:pruning-privacy}}

\begin{proof}
    First, auctioneer privacy. By definition of $\sim_{(P, A)}$, we have $\bm{\theta} \sim_{(P, A)} \bm{\theta}'$ is equivalent to $o_A(P,\bm{\theta}) = o_A(P,\bm{\theta}')$. Pruning leaves strategies unchanged and does not remove the on-path history $o_A(P,\bm{\theta})$, so $o_A(P,\bm{\theta}) = o_A(P,\bm{\theta}')$ holds if and only if $o_A(\tilde{P},\bm{\theta}) = o_A(\tilde{P},\bm{\theta}')$. However, $o_A(\tilde{P},\bm{\theta}) = o_A(\tilde{P},\bm{\theta}')$ is in turn equivalent to $\bm{\theta} \sim_{(\tilde{P}, A)} \bm{\theta}'$. Thus $\bm{\theta} \sim_{(P, A)} \bm{\theta}'$ if and only if $\bm{\theta} \sim_{(\tilde{P}, A)} \bm{\theta}'$.

    Next, bidder privacy. Suppose $\bm{\theta} \sim_{(P, i)} \bm{\theta}'$. This happens if and only if:
    \begin{enumerate}
        \item $\theta_i = \theta'_i$;
        \item $\phi(\bm{\theta}) \sim_{\Omega_i} \phi(\bm{\theta'})$;
        \item bidder $i$ observes the same sequence of information sets and takes the same actions along the path of play of $\bm{\theta}$ and $\bm{\theta}'$ under protocol $P$.
    \end{enumerate}
    Since conditions one and two do not depend on the game $G$ or $\tilde{G}$ being played, it suffices to focus on condition three. Let $\{h_k(\bm{\theta})\}_k$ be the sequence of histories at type profile $\bm{\theta}$ and let $\{h_k(\bm{\theta})\}_{k: i(h_k) = i}$ be the subsequence of histories for which bidder $i$ is called to play. Then, condition three can be stated as
    \begin{equation}\label{pre-pruning}
        \{I(h_k(\bm{\theta}))\}_{k: i(h_k) = i} = \{I(h_k(\bm{\theta}'))\}_{k: i(h_k) = i}.
    \end{equation}
    For any on-path histories $h,h' \in \tilde{H}$, we have $I(h) = I(h') \Leftrightarrow \tilde{I}(h)=\tilde{I}(h')$, as $\tilde{I}$ is simply a restriction of $I$ to on-path histories. Thus, (\ref{pre-pruning}) is equivalent to 
    \begin{equation}\label{post-pruning}
        \{\tilde{I}(h_k(\bm{\theta}))\}_{k: i(h_k) = i} = \{\tilde{I}(h_k(\bm{\theta}'))\}_{k: i(h_k) = i}
    \end{equation}
    which is then equivalent to bidder $i$ observing the same sequence of information sets along the path of play of $\bm{\theta}$ and $\bm{\theta}'$ under $\tilde{P}$. Since the strategies $S$ are unchanged under pruning, the actions bidder $i$ plays are similarly unchanged. Thus, if bidder $i$ observes the same sequence of information sets and takes the same actions along the path of play of $\bm{\theta}$ and $\bm{\theta}'$ under protocol $P$, they will also observe the same sequence of information sets and take the same actions along the path of play of $\bm{\theta}$ and $\bm{\theta}'$ under protocol $\tilde{P}$.
\end{proof}

\subsubsection*{PROOF OF PROPOSITION \ref{prop:pruning-IC}}

\begin{proof}
    Suppose $\phi$ is dominant-strategy incentive compatible. Towards a contradiction, suppose $P$ is not ex-post incentive compatible. Then, there exists $i, \theta_i, \bm{\theta}_{-i}, \sigma_i'$ such that 
    \begin{equation}\label{protocol-not-DSIC}
        u_i(g(S_i(\theta_i), \bm S_{-i}(\bm{\theta}_{-i})), \theta_i) < u_i(g(\sigma_i', \bm S_{-i}(\bm{\theta}_{-i})), \theta_i).
    \end{equation}
    As $G$ is pruned, the terminal history reached under $(\sigma_i', \bm S_{-i}(\bm{\theta}_{-i}))$ must have been reached under $\bm{S}(\theta_i', \bm{\theta}_{-i})$ for some $\theta_i' \in \Theta_i$. Because $P$ implements $\phi$,
    $$g(S_i(\theta_i), \bm S_{-i}(\bm{\theta}_{-i})) = \phi(\theta_i, \bm{\theta}_{-i}) \text{ and } g(\sigma_i', \bm S_{-i}(\bm{\theta}_{-i})) = g(S_i(\theta_i'), \bm S_{-i}(\bm{\theta}_{-i})) = \phi(\theta_i', \bm{\theta}_{-i}).$$
    Plugging this back into (\ref{protocol-not-DSIC}) gives
    $$u_i(\phi(\theta_i, \bm{\theta}_{-i}), \theta_i) < u_i(\phi(\theta_i', \bm{\theta}_{-i}), \theta_i)$$
    which contradicts $\phi$ being dominant-strategy incentive compatible. 
\end{proof}

\end{document}